\documentstyle[epsfig,natbib2,natbibmnfix]{mn}

\newcommand{\be}{\begin{equation}}
\newcommand{\ee}{\end{equation}}
\newcommand{\bea}{\begin{eqnarray}}
\newcommand{\eea}{\end{eqnarray}}
\newcommand{\bc}{\begin{center}}
\newcommand{\ec}{\end{center}}

\def\gsim{ \lower .75ex \hbox{$\sim$} \llap{\raise .27ex \hbox{$>$}} }
\def\lsim{ \lower .75ex \hbox{$\sim$} \llap{\raise .27ex \hbox{$<$}} }

\renewcommand{\thefootnote}{\fnsymbol{footnote}}

\title{Simulations of cosmic ray feedback by AGN in galaxy clusters}

\author[Sijacki et al.]
       {\parbox{18cm}{Debora~Sijacki$^{1,2}$\footnotemark[1],
       Christoph~Pfrommer$^{3}$, Volker~Springel$^{1}$ and Torsten~A.
       En{\ss}lin$^{1}$}\vspace{0.3cm}\\ 
       $^1$Max-Planck-Institut f\"{u}r Astrophysik,
       Karl-Schwarzschild-Stra\ss{}e 1, 85740 Garching bei
       M\"{u}nchen, Germany\\ $^2$ Institute of Astronomy, Madingley
       Road, Cambridge, CB3 0HA, United Kingdom \\$^3$Canadian
       Institute for Theoretical Astrophysics, University of Toronto,
       60 St. George Street, Toronto, Ontario, M5S 3H8, Canada}

\begin{document}

\maketitle
\begin{abstract} 
Feedback processes by active galactic nuclei (AGN) appear to be a key
for understanding the nature of the very X--ray luminous cool cores
found in many clusters of galaxies. We investigate a numerical model
for AGN feedback where for the first time a relativistic particle
population in AGN-inflated bubbles is followed within a full
cosmological context. In our high-resolution simulations of galaxy
cluster formation, we assume that BH accretion is accompanied by
energy feedback that occurs in two different modes, depending on the
accretion rate itself. At high accretion rates, a small fraction of
the radiated energy is coupled thermally to the gas surrounding the
quasar, while in a low accretion state, mechanically efficient
feedback in the form of hot, buoyant bubbles that are inflated by
radio activity is considered. Unlike in previous work, we inject a
non-thermal particle population of relativistic protons into the AGN
bubbles, instead of adopting a purely thermal heating. We then follow
the subsequent evolution of the cosmic ray (CR) plasma inside the
bubbles, considering both its hydrodynamical interactions and
dissipation processes relevant for the CR population. This permits us
to analyze how CR bubbles impact the surrounding intracluster medium,
and in particular, how this contrasts with the purely thermal case. Due
to the different buoyancy of relativistic plasma and the comparatively
long CR dissipation timescale we find substantial changes in the
evolution of clusters as a result of CR feedback. In particular, the
non-thermal population can provide significant pressure support in
central cluster regions at low thermal temperatures, providing a
natural explanation for the decreasing temperature profiles found in
cool core clusters. At the same time, the morphologies of the bubbles
and of the induced X-ray cavities show a striking similarity to
observational findings. AGN feedback with CRs also proves efficient
in regulating cluster cooling flows so that the total baryon fraction
in stars becomes limited to realistic values of the order of $\sim 10\%$, more
than a factor of 3 reduction compared with cosmological simulations
that only consider radiative cooling and supernova feedback. We find
that the partial CR support of the intracluster gas also affects the
expected signal of the thermal Sunyaev-Zel'dovich effect, with typical
modifications of the integrated Compton-$y$ parameter within the
virial radius of the order of $\sim 10\%$.

\end{abstract}


\section{Introduction}

\renewcommand{\thefootnote}{\fnsymbol{footnote}}
\footnotetext[1]{E-mail: deboras@mpa-garching.mpg.de }

There is a growing body of observational evidence indicating that many
galaxy clusters harbor in their centre supermassive black holes (BHs),
which interact in a complex way with their surroundings. Recently,
spectacular images from the XMM-Newton and Chandra X-ray telescopes
\citep[e.g.][]{McNamara2005, Fabian2006, Forman2006} have shed some
light on this intricate interaction, showing clear imprints of central
active galactic nuclei (AGN) in the intracluster medium of their host
clusters. In particular, by comparing observations obtained at radio
wavelengths with X--ray images, it is often seen that there are X--ray
depressions in the innermost cluster regions which spatially correspond
to significant radio emission \citep[e.g.][]{Owen2000, Blanton2001,
Clarke2005}. The most plausible candidate for inducing these features is
the central BH, which is generating jet-inflated radio lobes, often
simply dubbed `bubbles'.

There has been considerable theoretical effort to understand the
interplay of AGN-driven bubbles with the intracluster medium (ICM)
\citep[e.g.][]{Churazov2001, Quilis2001, Brueggen2002, Ruszkowski2002,
DallaVecchia2004, Omma2004a, Sijacki2006a, Sijacki2006b,
Sijacki2007}. Most of these studies have, for simplicity, considered
bubbles that are filled with hot thermal gas, and by following their
evolution with time, they tried to constrain how much heat AGN bubbles
can deliver to the surrounding medium. However, the synchrotron and
inverse Compton emission detected from the radio lobes suggests that
they contain a significant amount of relativistic electrons and that
they are permeated with magnetic fields. Besides relativistic
electrons and magnetic fields, it is also very plausible that an
important part of the pressure support in the bubbles stems from
relativistic protons, especially if the AGN jets are heavy. To date
it is however not clear, both from an observational and a theoretical
point of view, what the precise composition of the bubbles, the
relative mixture of hot gas, and non-thermal particle components really
is. Also, we have no detailed knowledge about the strength and
configuration of the magnetic fields within the bubbles.

Regarding the contribution from hot thermal gas, the work of
\citet{Mazzotta2002} suggested that the bubble in the MKW 3s galaxy
cluster may have a predominantly thermal origin, while
\citet{Schmidt2002} found that in the Perseus cluster a thermal gas
component cooler than $\sim 11\,{\rm keV}$ filling the bubbles seems
ruled out. Moreover, it is not clear whether entrainment of thermal
gas by the radio jet is effective and how this should depend on the BH
properties. A study of a larger sample of galaxy clusters with central
X-ray cavities by \citet{Dunn2005} hints towards a rather complex
picture, leaving several possibilities still open, both for
significant pressure support by non-thermal protons or by hot thermal
plasma, and for scenarios where the magnetic fields are filamentary
at least in some of the cases.
 
Bearing in mind the outlined uncertainties it is still highly
interesting to investigate the possible effects of cosmic rays (CRs)
in AGN-inflated bubbles on cool core clusters, as considered already
by several authors \citep[e.g.][]{Boehringer1988, Loewenstein1991,
Mathews2007, Guo2007, Sanders2007, Ruszkowski2007}. These studies have highlighted different
channels of CR interaction with the thermal intracluster gas and have
demonstrated that CRs can be dynamically important in galaxy clusters,
contributing up to $50\%$ of the central cluster pressure (but see
  \citet{Churazov2007} for somewhat lower estimates of CR pressure support of
  order of $10-20\%$). Nevertheless,
due to the complicated nature of CR physics,
most of these works were based on analytical or 1D numerical
calculations, and neglected any cosmological evolution of the host
clusters, and hence any possible dependence on its dynamical
state. On the other hand, cosmological simulations of CRs produced at
structure formation shocks by, for example, 
\citet{Miniati2001, Miniati2002, Ryu2003, Ryu2004} have highlighted
the importance of a realistic cosmological setting for a more fateful
generation, distribution and following evolution of CRs. 

Recently, \citet{Ensslin2007} have proposed a simplified formalism for
the treatment of CR protons that it suitable for implementation and
use in self-consistent cosmological codes, as subsequently
demonstrated by \citet{Jubelgas2007} and \citet{Pfrommer2006}. In
these numerical studies, the CR source processes considered were
restricted to supernovae (SNe) and structure formation shock
waves. While it was found that these CR sources can affect the
interstellar medium of low mass galaxies significantly, they did not
prevent the overcooling in the centres of galaxy groups and
clusters. However, the possible impact of CR protons generated by
central AGN has not been explored in these works, and this forms the
primary objective of our current study.

This paper is organized as follows. In Section~\ref{Methodology} we
outline the methodology we have adopted to simulate CR bubbles and to
follow their cosmological evolution. In Section~\ref{Isolated}, we
present test runs performed for isolated halo simulations in order to
analyze our numerical model in detail, while the bulk of our results
from cosmological simulations of galaxy cluster formation is described
in Section~\ref{Cosmological}. Finally, we discuss our findings and draw
our conclusions in Section~\ref{Discussion}.

\section{Methodology} \label{Methodology}

In this study we use an improved version of the massively parallel
TreePM-SPH code {\small GADGET-2} \citep{Gadget2, Springel2001}. The
SPH formulation adopted in the code manifestly conserves both energy
and entropy even in the presence of fully adaptive smoothing lengths,
as implemented by \citet{SH2002}. In addition to the gravity of dark
matter and baryons, and to ordinary hydrodynamics, the code tracks
radiative cooling of an optically thin plasma of hydrogen and helium,
immersed in a time-varying, spatially uniform UV background \citep[as
in][]{Katz1996}. A subresolution multi-phase model for the ISM is used
to follows star formation and supernovae feedback processes
\citep{S&H2003}. Furthermore, we adopt the model for BH seeding and
growth that has been suggested by \citet{DiMatteo2005} and
\citet{Springel2005b}, modified however by the introduction of a
second mode of AGN feedback to model radio activity, as described by
\citet{Sijacki2007}. These two modes are thermal heating from quasars at
high BH accretion rates (BHARs), and mechanical feedback in the form
of hot, buoyant bubbles occurring at low accretion rates.

In this study, we combine the AGN feedback prescription at low BHARs
with a model for CR treatment that has been developed, implemented,
and discussed by \citet{Ensslin2007}, \citet{Jubelgas2007}, and
\citet{Pfrommer2006}. This extension of our model allows us to account
for a non-thermal component permeating the radio lobes. In this
section, we outline the most important features of our combined BH and
CR models, and we describe in detail the approximations we have
adopted to model non-thermal particle populations in AGN-driven
bubbles.
\vspace{1.truecm}
\subsection{Black hole model}

We represent BHs in the code by collisionless sink particles of
initially very small mass, and we allow them to grow via gas accretion
and through mergers with other BHs that happen to get sufficiently
close. During the growth of structure, we seed every new dark matter
halo above a certain mass threshold (e.g. $5 \times 10^{10}\,h^{-1}
{\rm M}_\odot$) with a central BH of mass $10^{5}\,h^{-1} {\rm
M}_\odot$, provided the halo does not contain any BH yet. The seeding
is accomplished on-the-fly by frequently invoking a parallel
friends-of-friends algorithm. Once seeded, BHs can then grow by local
gas accretion, with an accretion rate estimated with the
Bondi-Hoyle-Lyttleton formula \citep{Hoyle1939, Bondi1944,
Bondi1952}. We impose a limit equal to the Eddington rate on the
maximum allowed BHAR.

As for the BH feedback processes, we assume that they are composed of
two physically distinct modes, depending on the BHAR itself. At high
accretion rates, i.e.~above $10^{-2}$ in Eddington units, we assume
that the BH is in a radiatively efficient phase, where a small
fraction of the bolometric accretion luminosity is coupled thermally
to the local gas particles around the BH, with an efficiency of $5\%$
and a spherical injection kernel. Instead, at low BHARs, we
conjecture that the BH growth is characterized by a radiatively
inefficient accretion flow, where most of the feedback is in a
mechanical form, manifesting itself by hot buoyant bubbles that rise
through the intragroup or intracluster medium. We relate the bubble
energy content, radius and duty cycle with the BH physics, and adopt a
somewhat higher efficiency of mechanical feedback of $20\%$. We note
that \citet{Allen2006} recently showed that there exists a tight
correlation between the Bondi accretion rates calculated based on
observed gas temperature and gas density profiles and estimated BH
masses, and the actual power emerging from these systems in
relativistic jets. This lends observational support to our simple
estimates of the BHARs.

\subsection{CR implementation}

We represent the CR population in each gaseous fluid element by a
relativistic population of protons, which we approximately describe
with an isotropic power-law distribution function in momentum
space. In the simple formalism adopted here, this distribution
function is fully defined by the power law slope $\alpha$, its
normalization $C$, and a dimensionless low momentum cut-off $q$: ${\rm
d}n/{\rm d}p \,=\,Cp^{-\alpha} \theta(p-q)$. Here $\theta(x)$ denotes
Heaviside step function and the dimensionless momenta are expressed in units
of $m_{\rm p}\,c$. We assume for
clarity that the only source of CRs is given by the radiatively
inefficient accretion mode of BHs, while we neglect other
contributions from supernovae or particle acceleration at cosmic
structure formation shocks, which have been extensively discussed in
previous work \citep{Jubelgas2007, Pfrommer2006, Pfrommer2007a,
Pfrommer2007b, Pfrommer2007c}. We include CR loss processes in the
form of thermalization by Coulomb interactions and losses due to
hadronic interactions. We expect that within dilute bubbles these loss
mechanisms should play a subdominant role compared to the pressure
loss due to adiabatic expansion of the buoyantly rising
bubbles. Therefore, we believe that our simulations also are
representative, at least qualitatively, of the case of a radio plasma
that is dominated by a relativistic electron-positron population. 

Even though in the present implementation of CR processes in {\small
GADGET-2}, accounting for CR diffusion is in principle possible, in
this study we refrain from trying to model it because this treatment
of diffusion is only isotropic at present. Nevertheless, during the rise of
a bubble, CR diffusion should be significantly suppressed
perpendicular to the magnetic field lines that drape around the
bubbles, only allowing CRs to diffuse efficiently in the wake of the
bubbles \citep[see e.g.][]{Sanders2007, Ruszkowski2007}. Thus, isotropic CR
diffusion would be a rather poor representation of this process, and
neglecting diffusion altogether probably mimics the possible influence of
magnetic draping effects with higher realism. Note, however, that the
  impact of magnetic fields on CR diffusion out of the bubbles is still poorly
understood and is a matter of debate, hence a detailed treatment of this issue
is beyond the scope of this work.

\subsection{A model for CR bubbles} 

In the numerical framework we adopted for the description of CRs, we
need to determine the power law slope $\alpha$ of the momentum spectrum
representative for our system, its injection cut-off $q_{\rm inj}$, and
the fraction $f_{\rm CR}$ of the energy released by the BH that actually
goes into the CRs. Once these values have been chosen we can follow the
bubble's evolution and the loss processes of the CR component
self-consistently, and compare with the case where the AGN-driven
bubbles are purely thermal.

For the slope $\alpha$ we have tested values from $2.1$ to $2.4$,
where a steeper power-law slope corresponds to a distribution with
more low-energy CRs. These can thermalize faster and have a less
relativistic equation of state that is closer to the `harder' thermal
case. We note that this range of values for the spectral slope is
consistent with observational findings for the electron population in
FRI sources, especially considering young systems
\citep[e.g.][]{Birzan2004, Dunn2005}. Similarly to the case in which
CRs are produced by supernovae \citep{Jubelgas2007} we establish the
injection cut-off $q_{\rm inj}$ below which the CR energy is instantly
thermalized. Equating the injection and loss time scale, we solve for
the injection cut-off $q_{\rm inj}$ using equations $(18)$ and $(19)$
from \citet{Jubelgas2007} and by assuming an initial value of the
intrinsic injection cut-off $q_{\rm init} \sim 1$ for simplicity. Our
results do not depend on a particular choice for $q_{\rm init}$,
provided $q_{\rm init} \le 1$ since Coulomb losses rapidly remove the
low energy part of the CR spectrum, which gets almost instantly
thermalized.

Finally, at a fixed mechanical feedback efficiency, we can choose
whether to fill the AGN-inflated bubbles with relativistic gas
exclusively, or to allow some of the hot thermal gas to permeate them as
well, by regulating the parameter for the energy fraction going into
CRs, $f_{\rm CR}$. Note that, if Coulomb losses are important some
fraction of initial energy delivered by AGN will be thermalized
instantly as described above, such that $f_{\rm CR}$ can regulate
only the remaining fraction of energy that in principle is available
for injection into the CR population directly. So far there is no clear
consensus in the observational literature \citep[e.g.][]{Mazzotta2002,
  Schmidt2002, Sanders2007, Simionescu2007} whether there is a
significant thermal component inside radio bubbles or not, thus one of
our main aims is to understand to what extent the numerical
simulations predict noticeable changes in our results for different
values of $f_{\rm CR}$, which can provide interesting insights with
respect to this question.
 
\section{CR bubbles in isolated halo simulations} \label{Isolated}

In this section we focus on idealized simulations of isolated galaxy
clusters that harbour an AGN at their centre. The initial conditions for
the runs consist of a gaseous cluster atmosphere in hydrodynamic
equilibrium within a static NFW \citep{Navarro1996, Navarro1997} dark
matter halo, as described in more detail in \cite{Sijacki2006a}. These
systems are then evolved under the influence of radiative cooling, star
formation, BH growth and different types of AGN feedback. The simplicity
of these simulations relative to a full cosmological setting permits us
to unambiguously identify the specific signatures of AGN-inflated
bubbles that are filled with CRs, and to understand the differences that
occur with respect to purely thermal bubbles.

\subsection{Thermal versus CR content of bubbles}

We start our analysis by considering a $10^{15}h^{-1}{\rm M_\odot}$ halo
where in regular time intervals of $10^8\,{\rm yr}$ a pair of spherical
bubbles is injected into the innermost cluster region. We have fixed the
energy content of the bubbles to be $2.5\times 10^{60}{\rm erg}$ each,
and the bubble radius to be $30\, h^{-1}{\rm kpc}$. We have performed
two identical runs, with the only difference being that in the first run
the energy was injected as thermal energy into the bubbles, while in the
second run the bubbles were filled with CRs instead.

\begin{figure*} \centerline{ \vbox{ \hbox{
\psfig{file=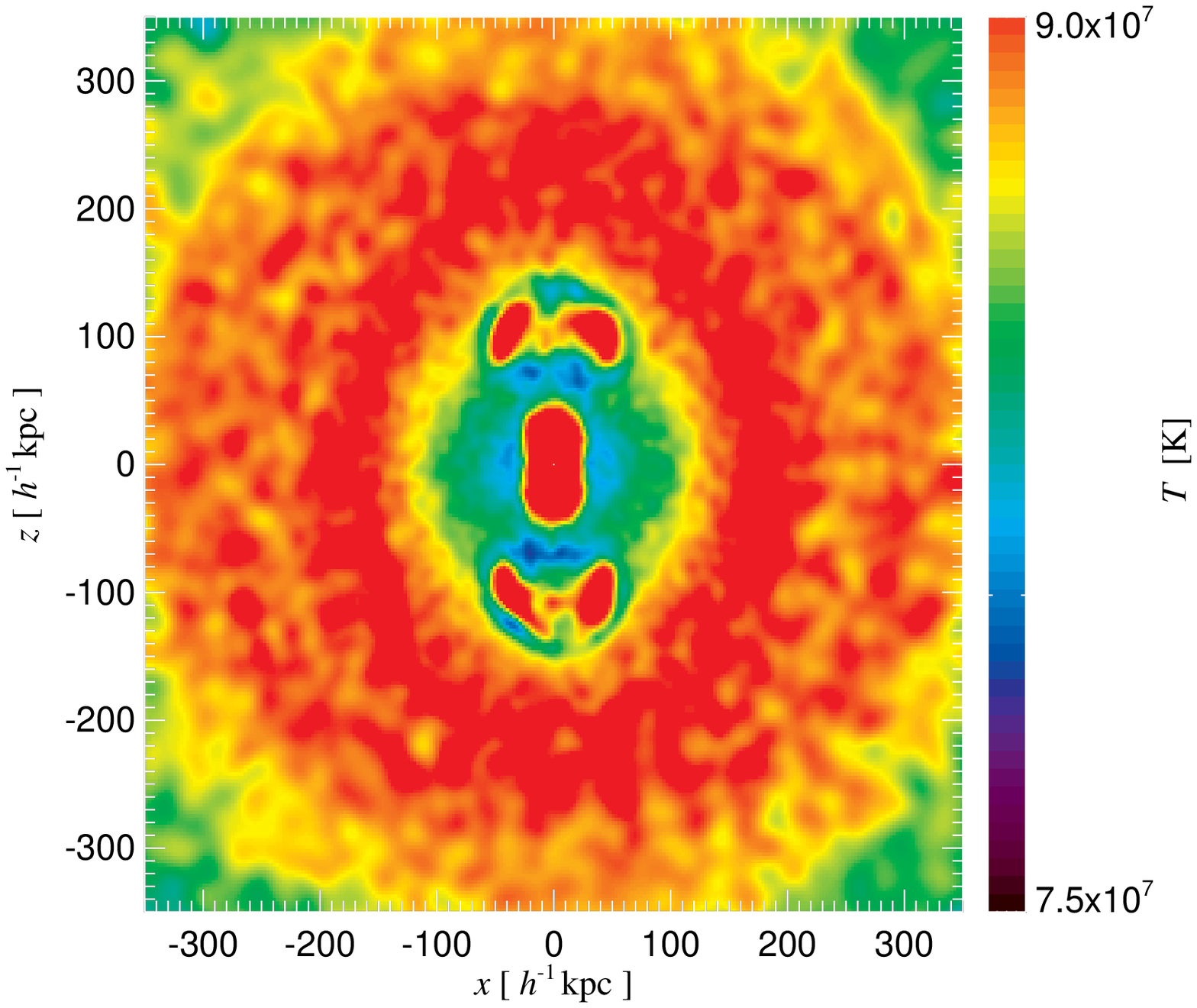,width=6.5truecm,height=5.8truecm}
\hspace{-0.4truecm}
\psfig{file=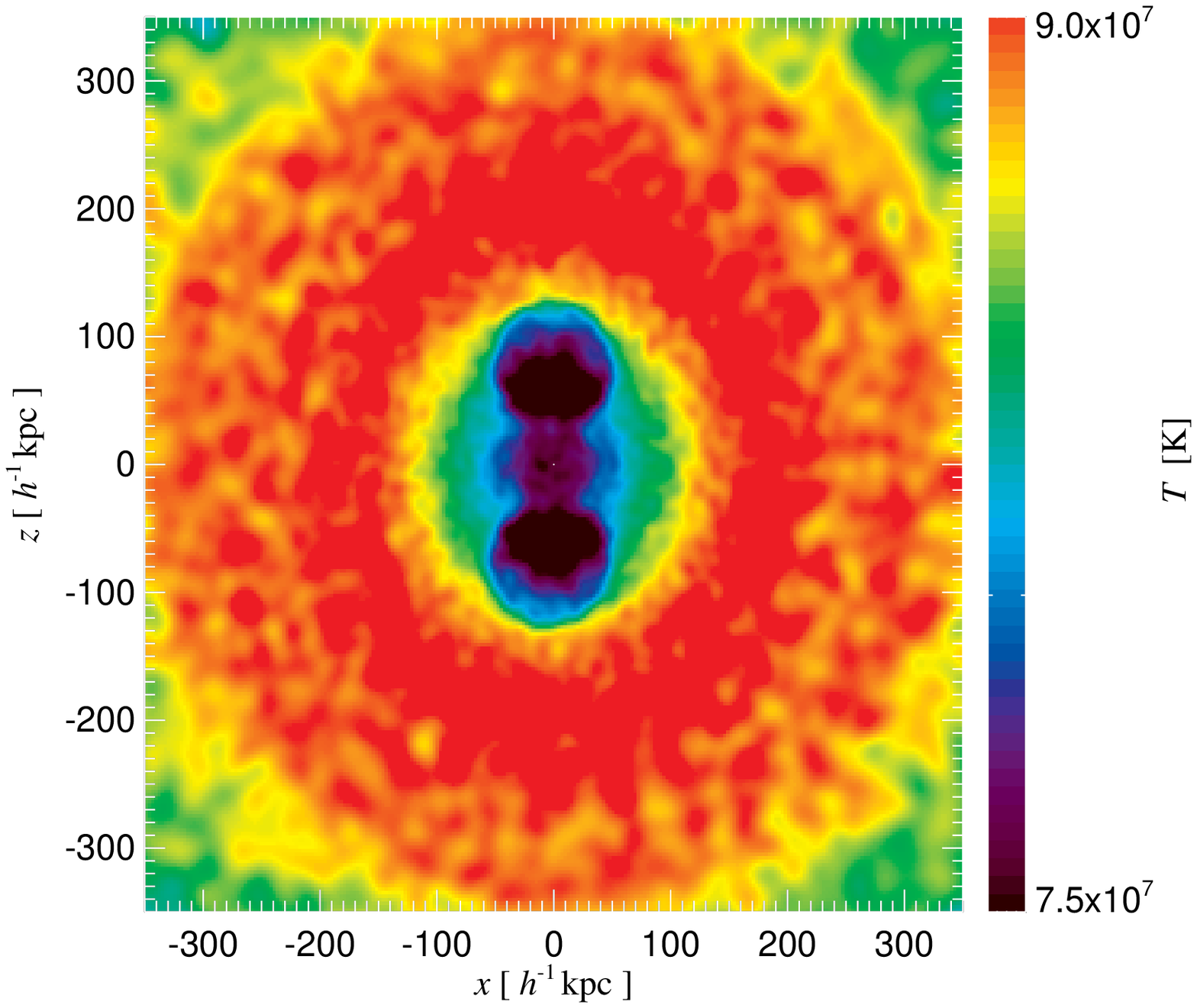,width=6.5truecm,height=5.8truecm}
\hspace{-0.4truecm}
\psfig{file=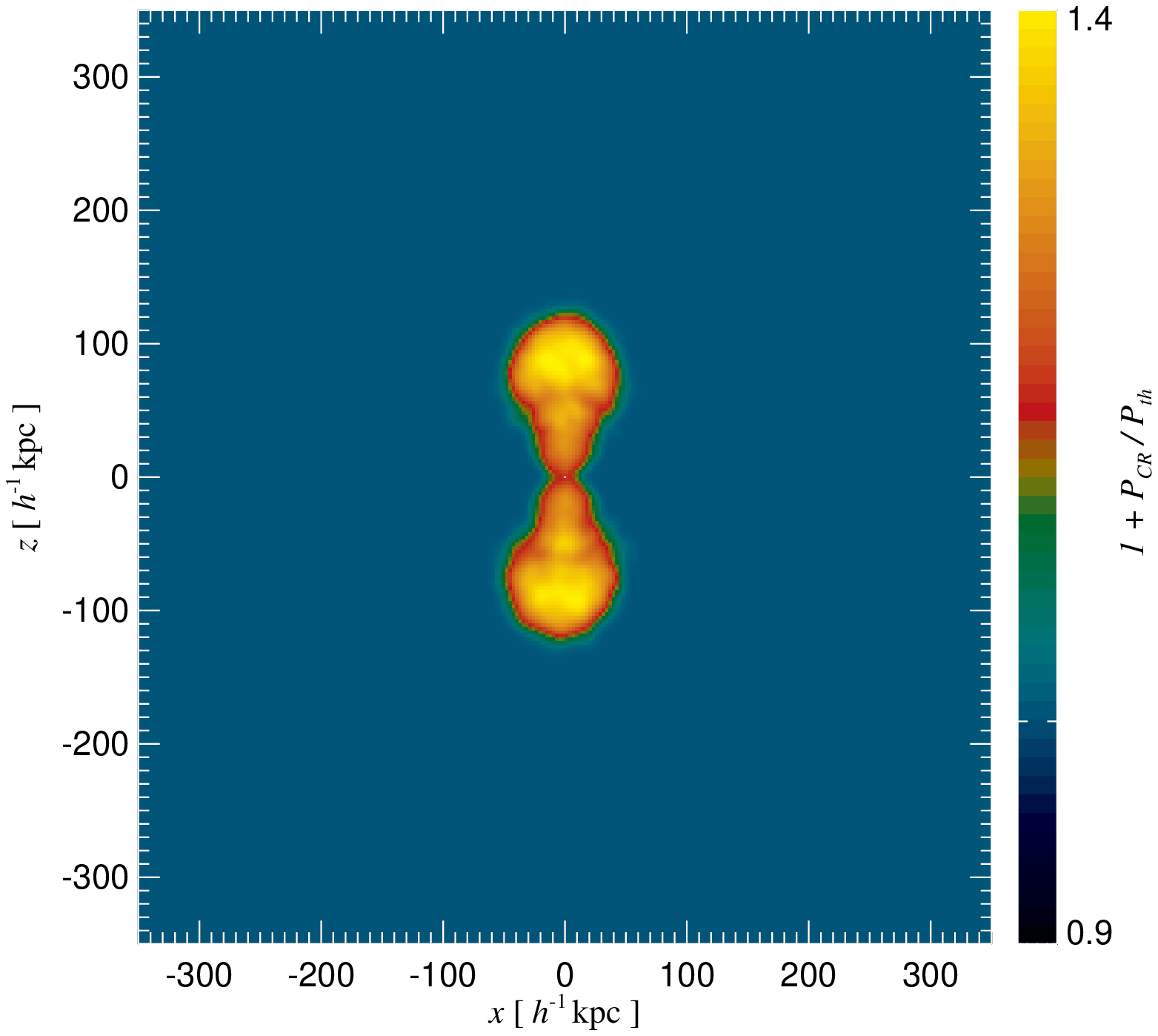,width=6.5truecm,height=5.8truecm}
} \vspace{-0.5truecm} \hbox{
\psfig{file=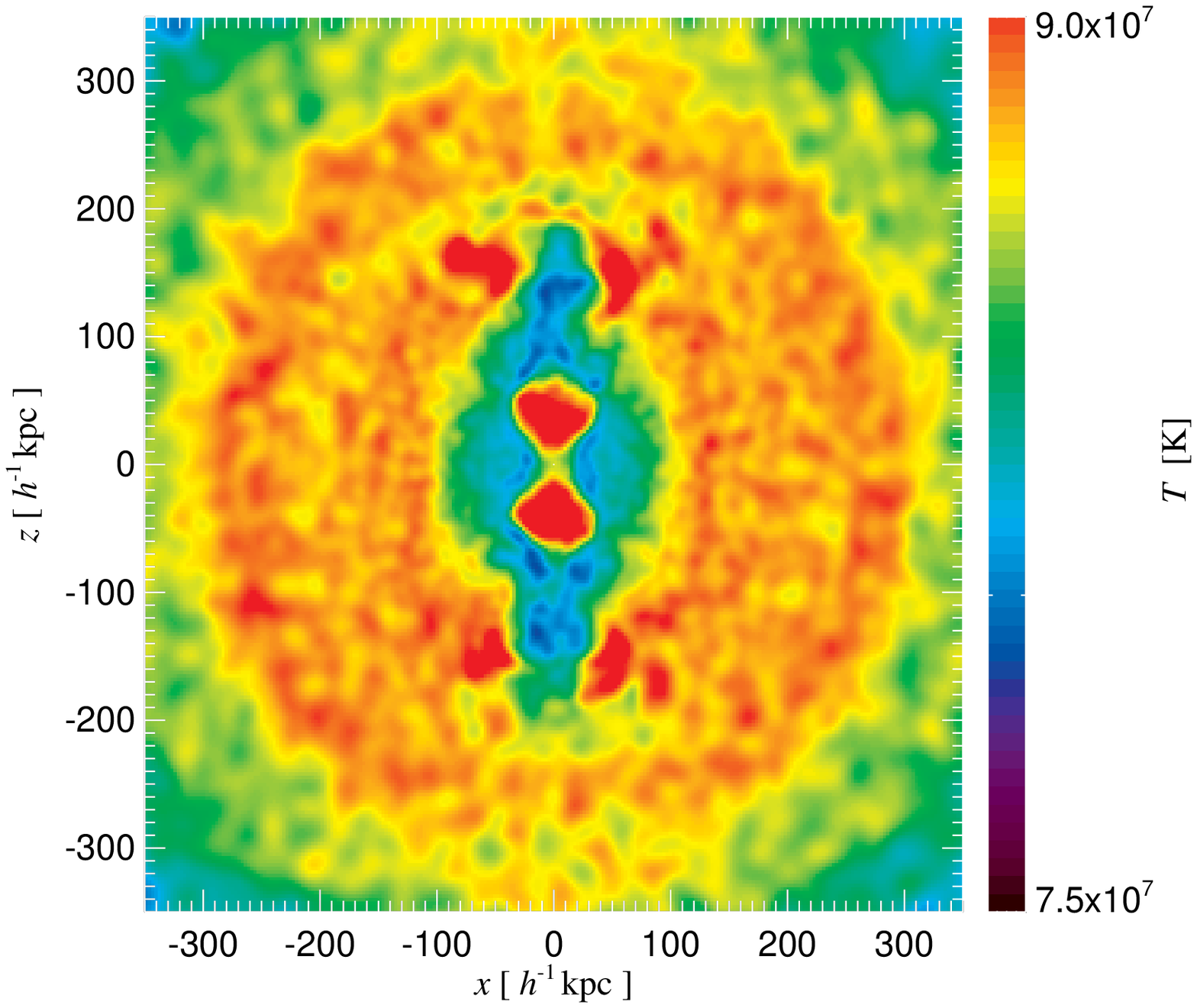,width=6.5truecm,height=5.8truecm}
\hspace{-0.4truecm}
\psfig{file=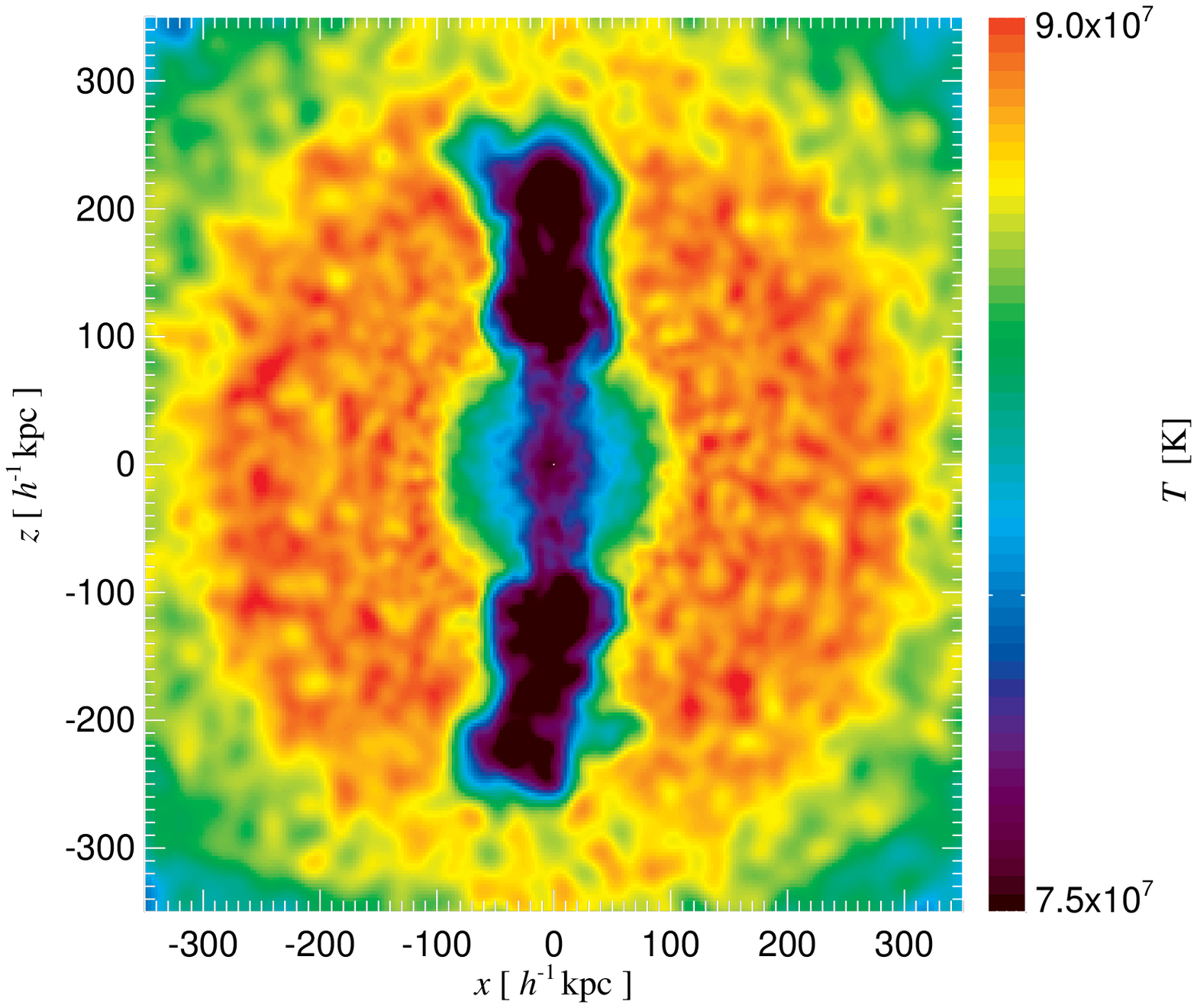,width=6.5truecm,height=5.8truecm}
\hspace{-0.4truecm}
\psfig{file=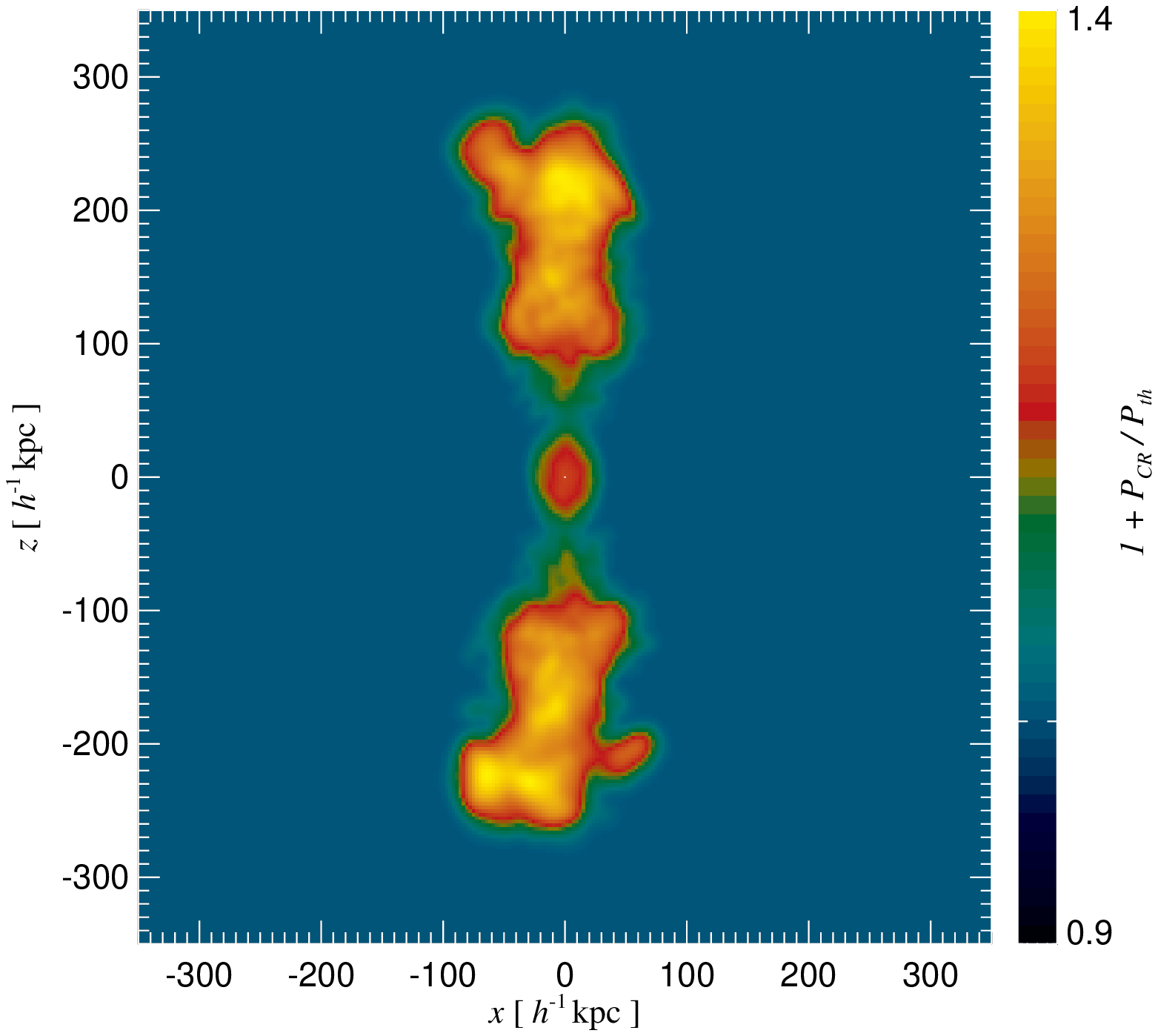,width=6.5truecm,height=5.8truecm}
} \vspace{-0.5truecm} \hbox{
\psfig{file=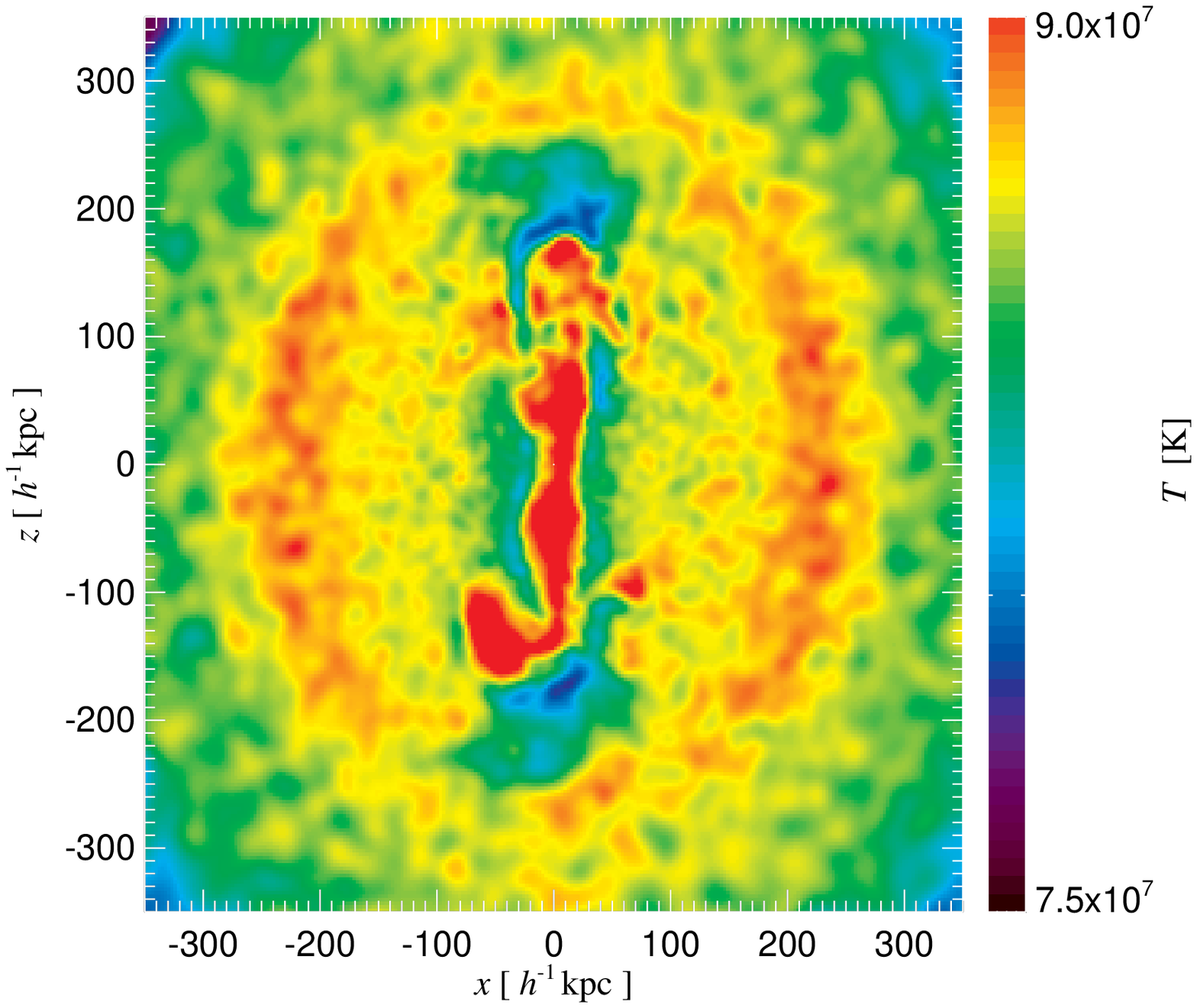,width=6.5truecm,height=5.8truecm}
\hspace{-0.4truecm}
\psfig{file=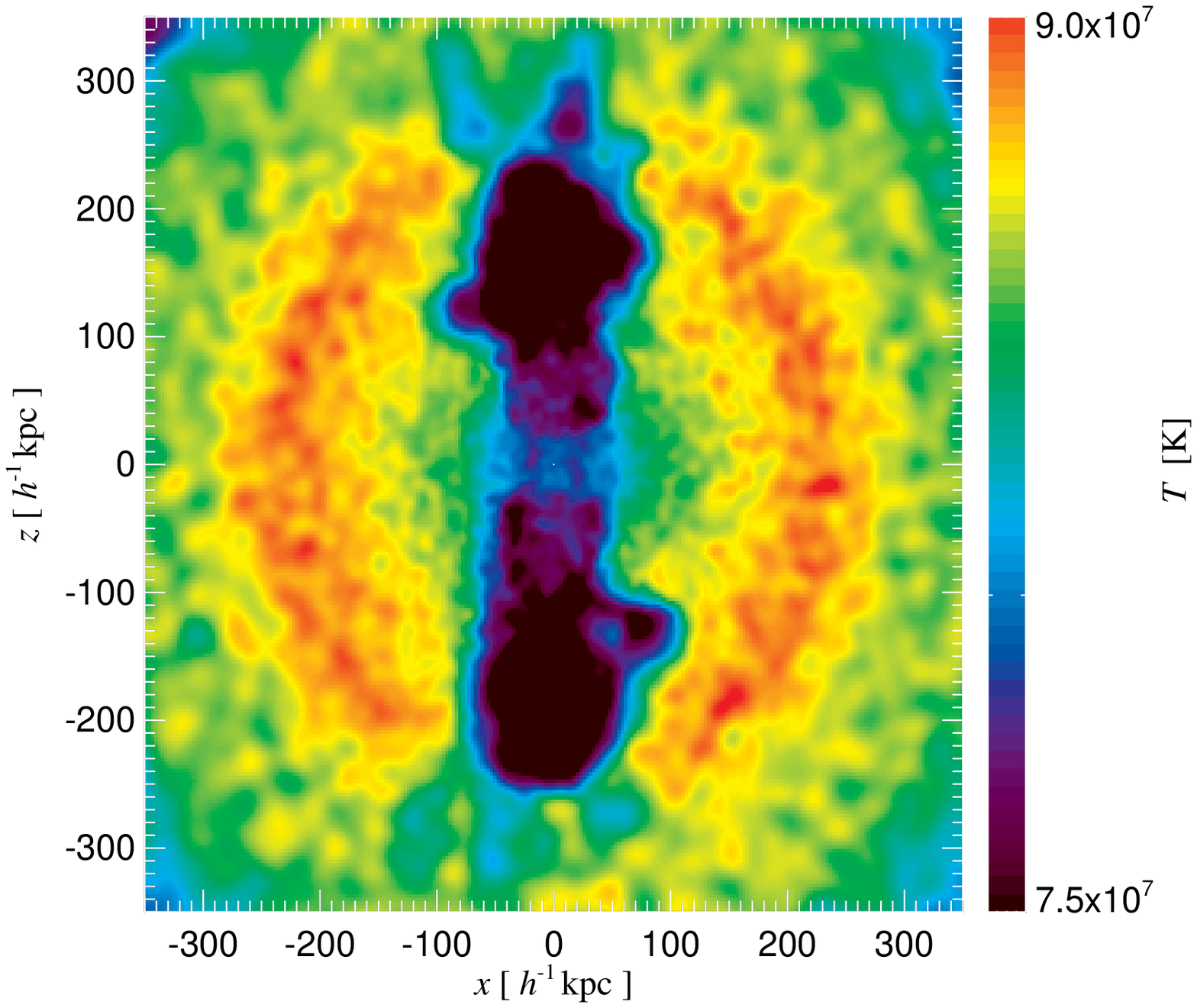,width=6.5truecm,height=5.8truecm}
\hspace{-0.4truecm}
\psfig{file=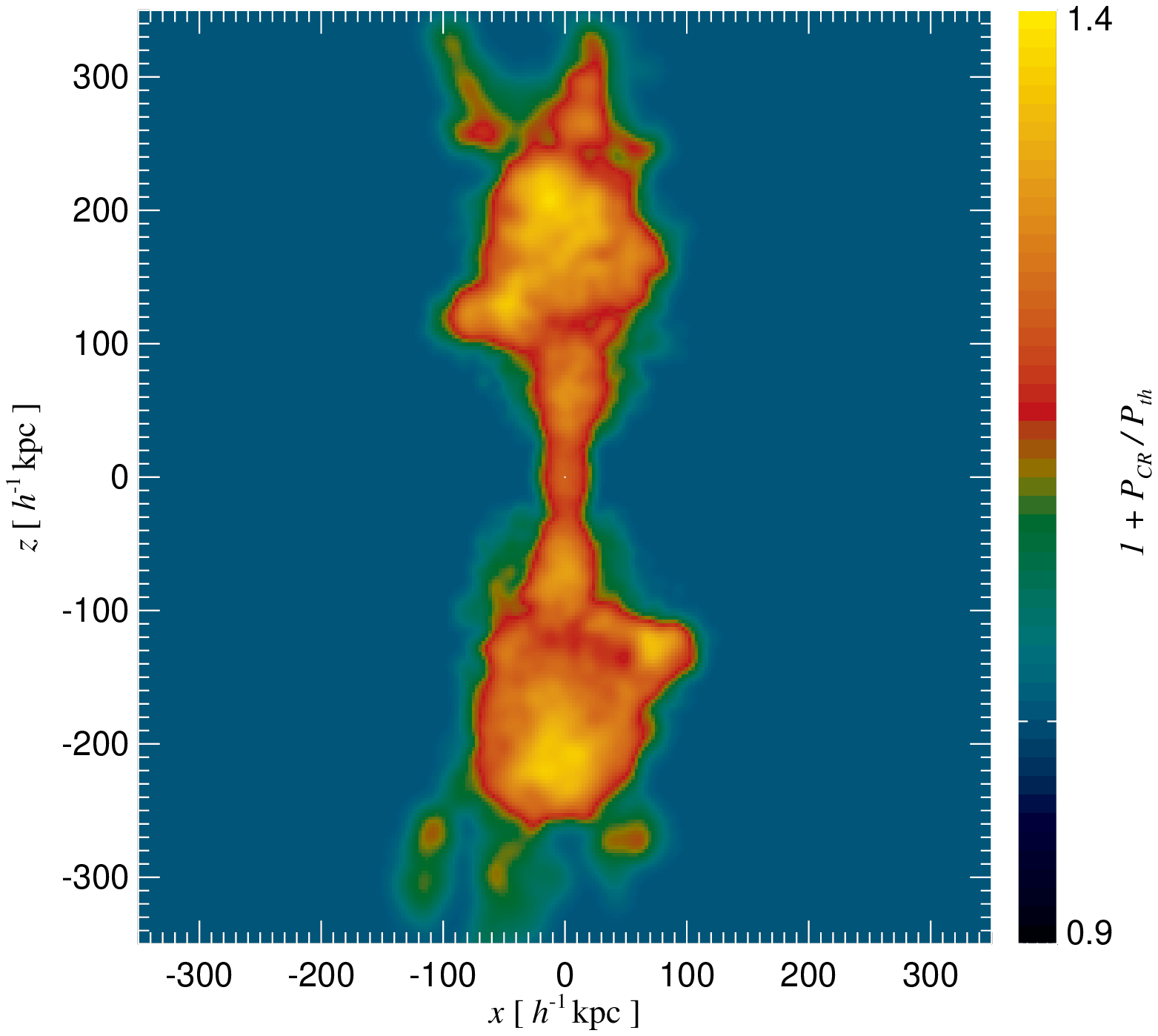,width=6.5truecm,height=5.8truecm}
} }}
\caption{Projected mass-weighted temperature and pressure maps of the
central cluster region at three different evolutionary stages, at times:
$0.07\,t_{\rm Hubble}$ (top panels), $0.12\,t_{\rm Hubble}$ (middle
panels), and $0.24\,t_{\rm Hubble}$ (bottom panels). The left-hand
panels show temperature maps in the run where thermal bubbles have been
injected in regular time intervals. The central panels illustrate a
simulation where the bubbles were initially filled with relativistic
particles instead. For this run, the regions where the CR pressure is
significant with respect to the thermal pressure are shown in the
right-hand panels. It can be clearly seen that the bubble dynamics,
coherence and maximum cluster-centric distance reached are affected by
the presence of a relativistic component filling the bubbles.}
\label{iso_maps}
\end{figure*}

In Figure~\ref{iso_maps}, the panels from top to bottom show a time
evolution sequence of this AGN-heated cluster. The panels on the
left-hand side are projected temperature maps of the central cluster
region for the case with thermal bubbles, while the panels of the middle
column show the corresponding region in the run where the bubbles are
filled with relativistic particles. To more clearly show the location of
the CR bubbles, the panels on the right-hand side give the ratio of the
projected CR pressure to the projected thermal pressure, so that regions
that are permeated with a significant CR component can be clearly seen.

It is evident from this figure that the CRs modify both the bubble
morphologies as well as their dynamics. The most striking difference in the
temperature maps is that the CR bubbles actually appear as X--ray cavities, as
has been observed in a number of clusters \citep[e.g.][]{Owen2000,
Blanton2001, Nulsen2005, McNamara2005, Forman2006, Fabian2006}. This is due to
the fact that once the bubbles expand rapidly, their pressure drops more
slowly than in the thermal case due to their softer equation of state, so that
they become larger and thermally cooler\footnote{Note that prior to the CR
injection into the bubble particles these have already by construction thermal
energy content comparable to the surrounding ICM particles to which we add
non-thermal energy released by the AGN. This thermal component inside the
bubbles is in most of the cases negligible with respect to the CR component.},
as can be seen by comparing in the maps the characteristic sizes of bubbles in
the cases with thermal and CR injection. Thus, even though the bubble radius
was the same at the moment of injection in both cases, the subsequent
evolution leads to a marked difference in the size evolution of the bubbles.

Furthermore, the CR bubbles remain coherent for a significantly longer
time than the thermal bubbles, and they reach larger cluster-centric
distances as well. Support for this finding can be seen in the middle
panels of Fig.~\ref{iso_maps}, where the thermal bubbles that were still
clearly defined in the top panel are already shredded when reaching a
distance of $\sim 150\, h^{-1}{\rm kpc}$, while the CR bubbles have
risen to $\sim 200\, h^{-1}{\rm kpc}$ and are still intact at the same
time. One reason for this behaviour lies in the significantly longer
cooling time of the relativistic gas inside the bubbles compared with
the thermal cooling time. Thus, while we injected initially the same
energy in both runs, the CR bubbles preserve their entropy content for a
longer time. Another reason lies in the enhanced buoyancy of the CR
bubbles thanks to their larger size.

\begin{figure*} \centerline{ \vbox{\hbox{ \hspace{-0.3truecm}
\psfig{file=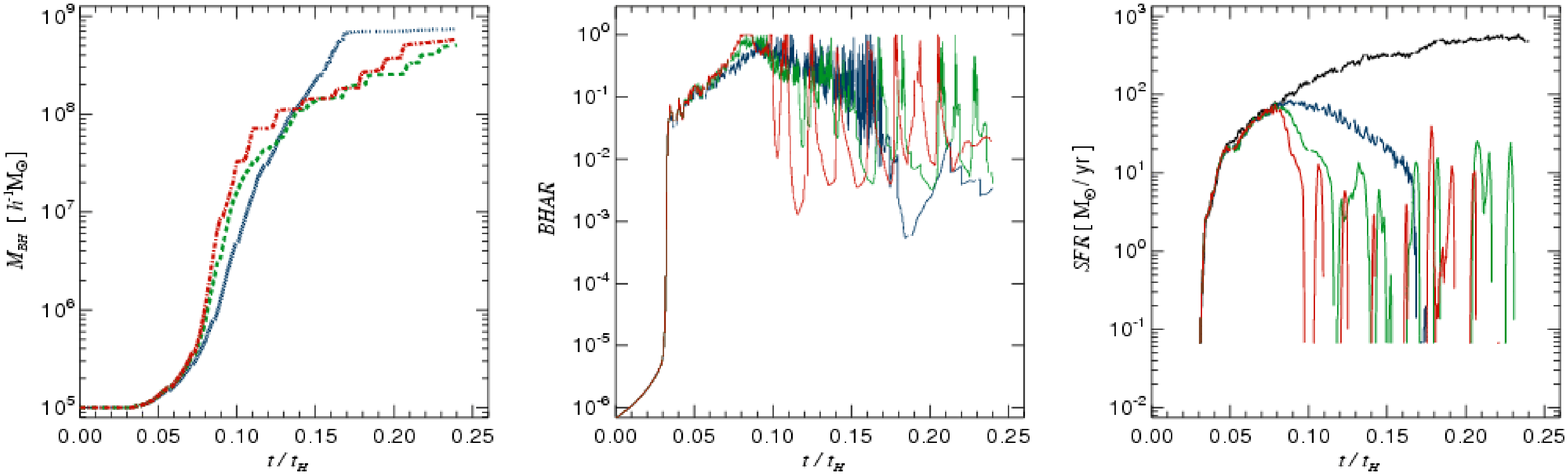,width=18.9truecm,height=5.8truecm }}
\hbox{ \psfig{file=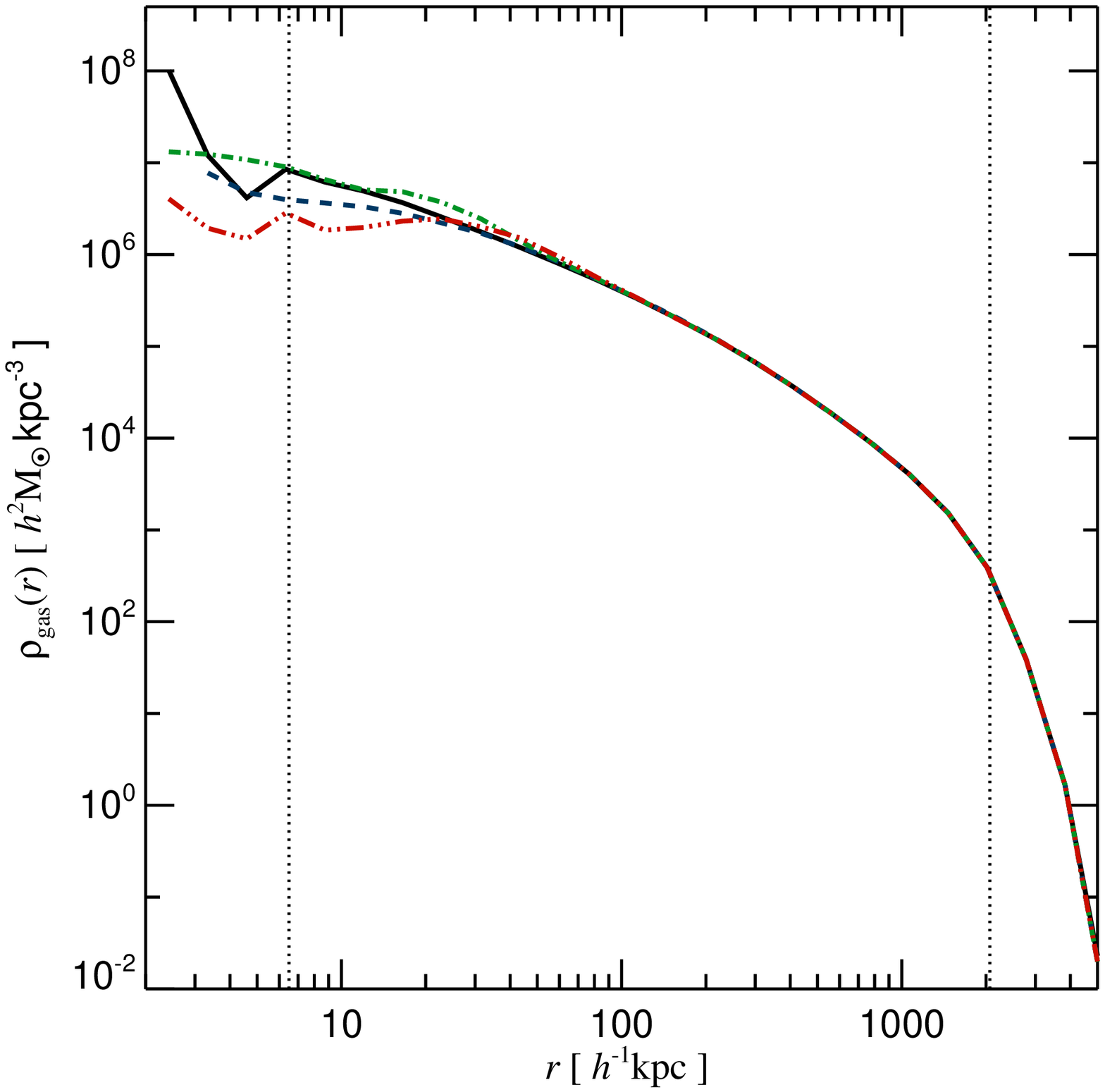,width=6.truecm,height=5.8truecm}
\hspace{-0.truecm}
\psfig{file=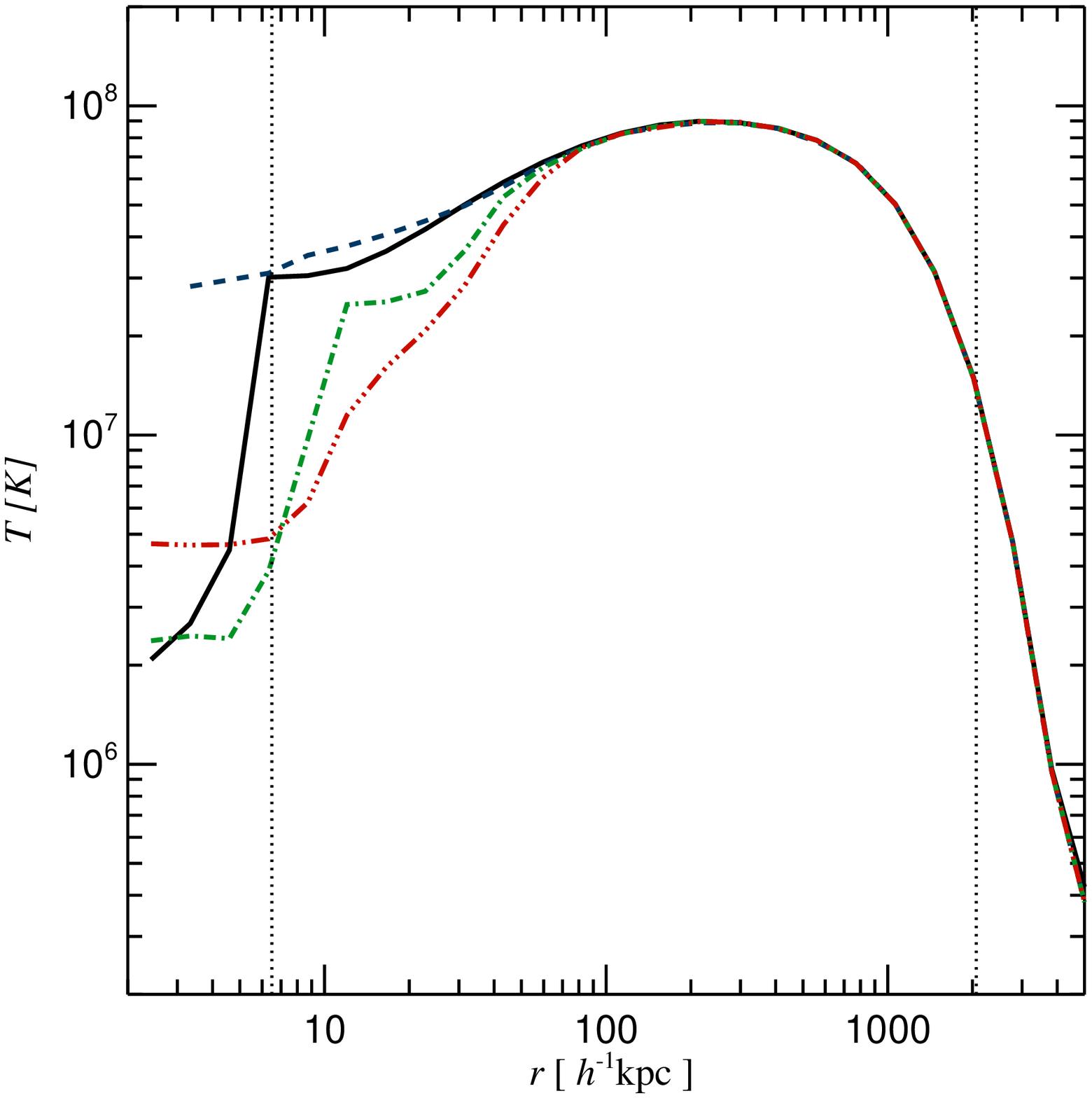,width=6.truecm,height=5.8truecm}
\hspace{-0.truecm}
\psfig{file=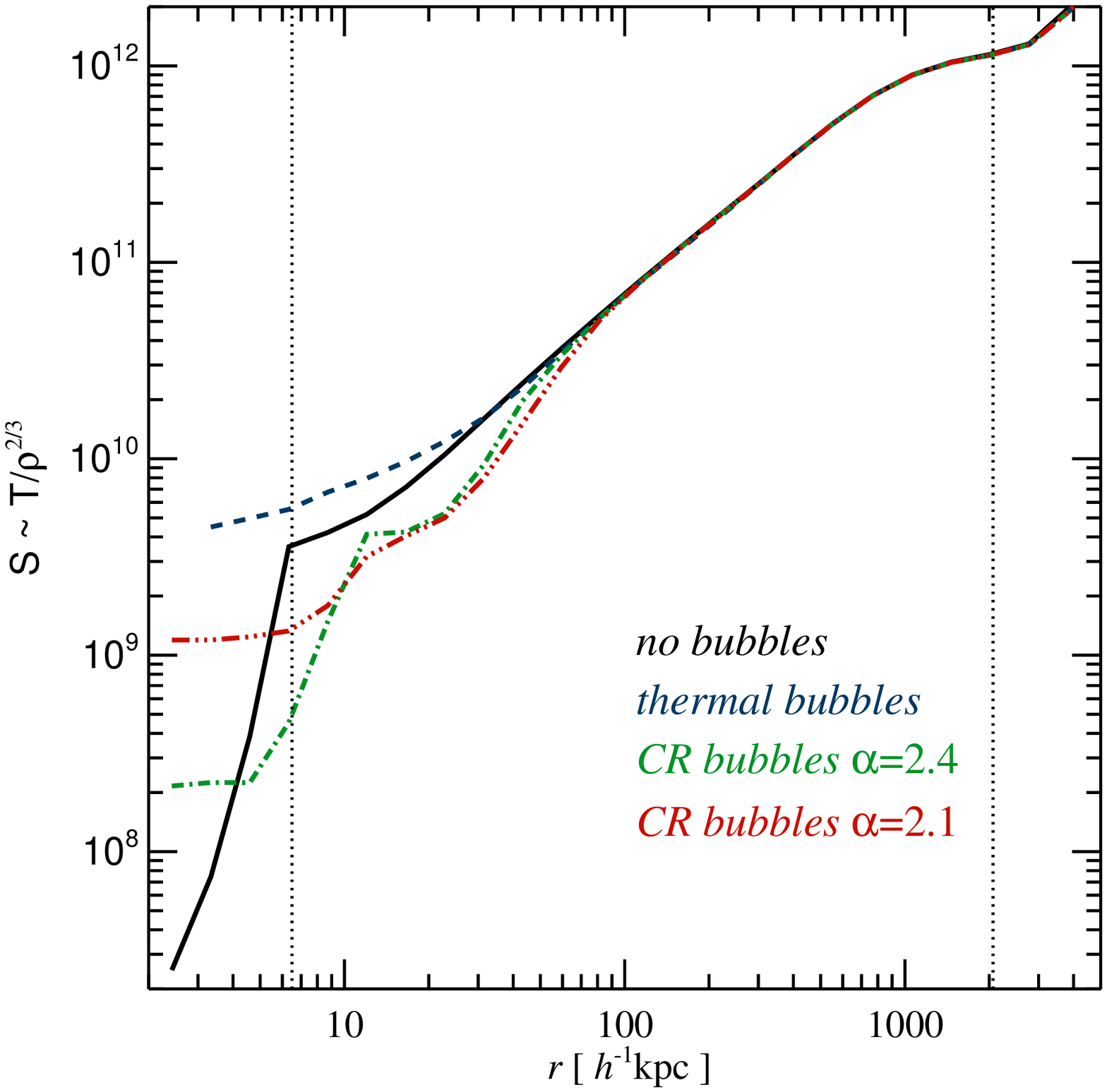,width=6.truecm,height=5.8truecm}}}}
\caption{The top panels show the BH mass growth, BHAR in Eddington
units, and star formation rate (SFR) as a function of time for four
different runs: black lines denote the case with cooling and star
formation only, blue lines show results where the AGN-driven bubbles are
thermal, while the green and red lines refer to the case of CR
bubbles. The run with a steeper spectral index of $\alpha=2.4$ is
represented with green lines, while the case of $\alpha=2.1$ is drawn
with red lines. The bottom panels show radial profiles of gas density,
mass-weighted temperature, and entropy, for the same set of runs and
with the same colour-coding. The vertical dotted lines denote the
gravitational softening length and the virial radius of this halo.}
\label{iso_profiles}
\end{figure*} 

\subsection{Self-regulated CR bubble heating}

Now we turn to a more detailed analysis that involves a growing BH. To
this end we introduce at the beginning of the simulations a seed BH of
small mass at the centre in order to see how feedback by CR bubbles
fares in establishing self-regulated AGN activity. These test runs are
analogous to the ones presented in \cite{Sijacki2007}, where a more
detailed description of the set-up can be found. We also test variations
of the CR power law slope from a value of $2.1$ to $2.4$, as illustrated
in Figure~\ref{iso_profiles}, but we fix $q_{\rm init} =1$ and $f_{\rm
CR}=1$.

First, we analyze the growth of the central BH as a function of time in
the top panels of Figure~\ref{iso_profiles}. It can be seen that the CR
bubbles lead to a self-regulated growth of the central BH similar to the
thermal case, but at the end of the simulated time-span, the BH reaches
a slightly smaller mass in the CR case. However, the detailed time
evolution in the case of CR bubbles is physically more complex than in
the thermal case. Initially, for $t < 0.1\,t_{\rm Hubble}$, the CR
bubbles are less effective in heating the ICM, and consequently the
central BH is growing more efficiently. During these early stages of BH
growth, the average CR pressure in the cluster central region is smaller
than the central pressure in the runs with thermal bubbles. Hence,
during this early phase the combined effect of Coulomb and hadronic
losses as well as the higher compressibility of composite CR plus
thermal gas leads to a less efficient CR bubble feedback, as will be
discussed in more detail in Section~\ref{Importance}. However, even
though this allows the gas to cool somewhat more efficiently towards the
centre the amount of stars formed is roughly the same in the thermal and
relativistic cases.

For $t > 0.1\,t_{\rm Hubble}$, the star formation rate begins to be more
suppressed in the run with CR bubbles, because at this stage $P_{\rm
CR}$ starts to be comparable to the local $P_{\rm th}$. In fact, for the
run with $\alpha = 2.1$, this transition occurs somewhat earlier than in
the simulation with a steeper $\alpha$, given that for $\alpha = 2.1$
the cosmic ray pressure reaches significant values sooner. It can be
seen that for $t > 0.1t_{\rm Hubble}$ the BHAR starts to oscillate
dramatically as a consequence of the presence of a dominant relativistic
particle component. A similar behaviour has been observed in the case of
star formation in small dwarf galaxies in the work by
\cite{Jubelgas2007}. When a bubble is injected into the ICM and its
$P_{\rm CR}$ is comparable or higher than the local $P_{\rm th}$ it will
provide an extra pressure support to the ICM. While rising buoyantly
towards the cluster outskirts and heating the surrounding gas, this
dramatically reduces the amount of gas that can be accreted by the
central BH. Successively, CR pressure will get dissipated and gas will
start to cool again leading to higher BHAR values, until another bubble
is triggered, and the cycle is repeated. Note that at late times due to
this behaviour of CR bubbles there is some residual star formation
activity and the BH growth is not as efficiently terminated as it is the
case for thermal bubbles.

The bottom panels of Figure~\ref{iso_profiles} show how the profiles of
gas density, temperature and entropy are affected by CR bubbles. The
radial temperature profile is always decreasing towards the central
regions in the runs with CR bubbles, reflecting the fact that a
significant fraction of non-thermal pressure builds up there. However,
the gas density can either be increased or reduced relative to the
thermal case, depending on whether $P_{\rm CR}$ is higher than $P_{\rm
th}$ at a given instant. Finally, the gas entropy is also somewhat
reduced in the innermost regions, mainly being driven by the variations
in temperature. Overall, the simulation with a flatter CR spectrum shows
a more significant imprint of CRs on the ICM, as expected due to the
more pronounced high-energy tail of the CR momentum distribution, so
that a significant level of $P_{\rm CR}$ can be maintained for a longer
time interval.

\begin{figure*} \centerline{ \vbox{\hbox{
\psfig{file=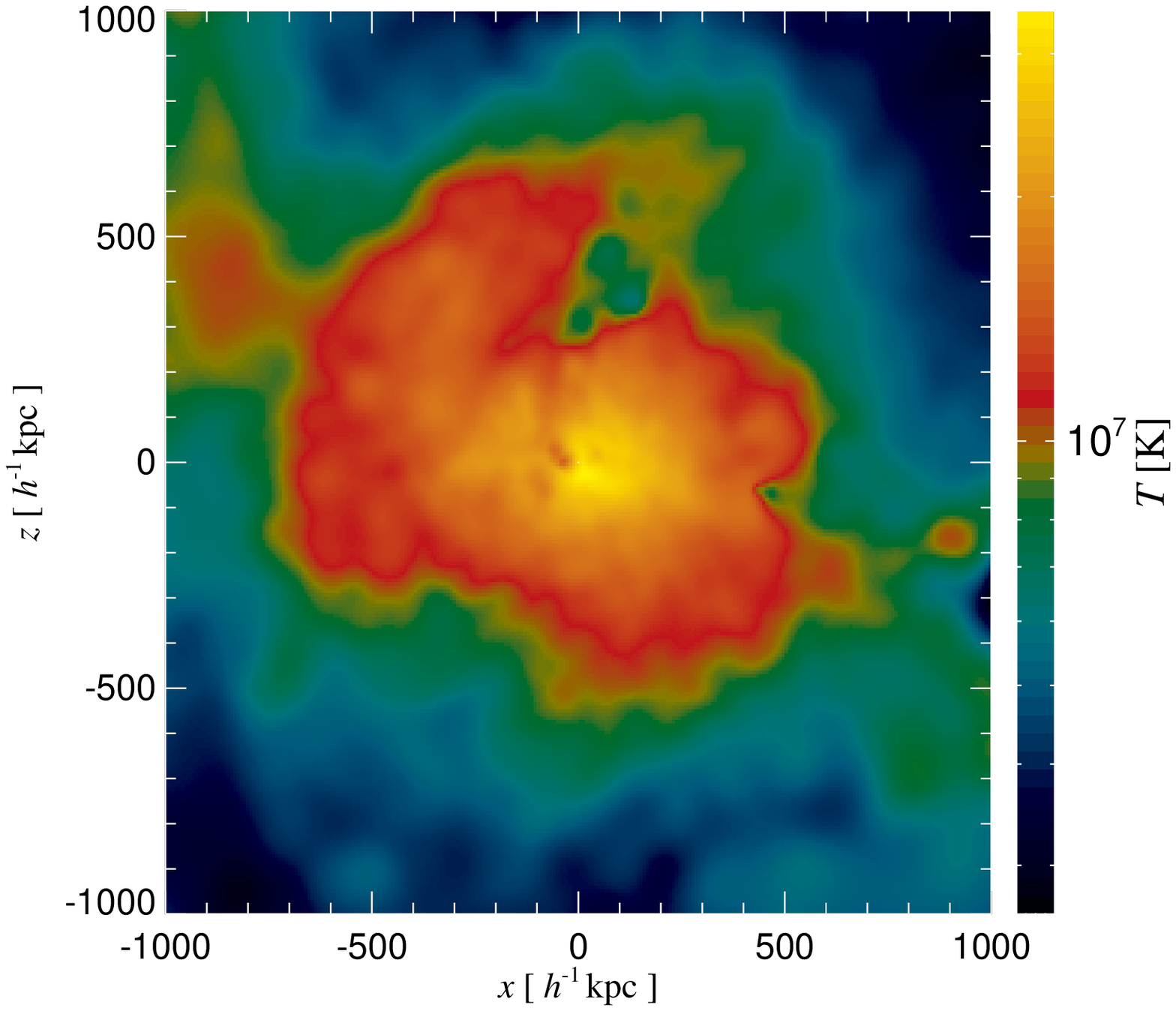,width=9.truecm,height=8.3truecm}
\hspace{-0.3truecm}
\psfig{file=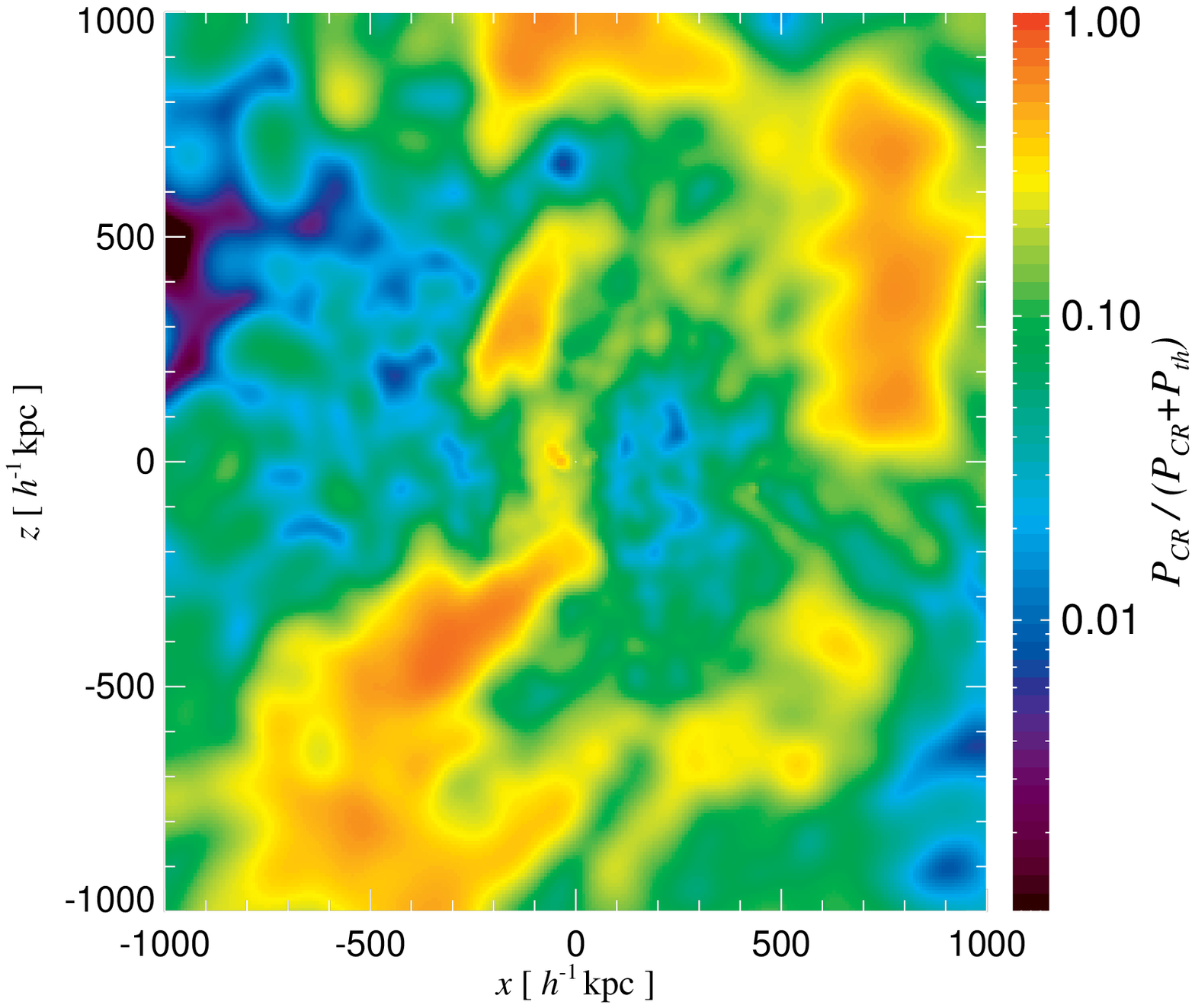,width=9.truecm,height=8.3truecm}}
\hbox{
\psfig{file=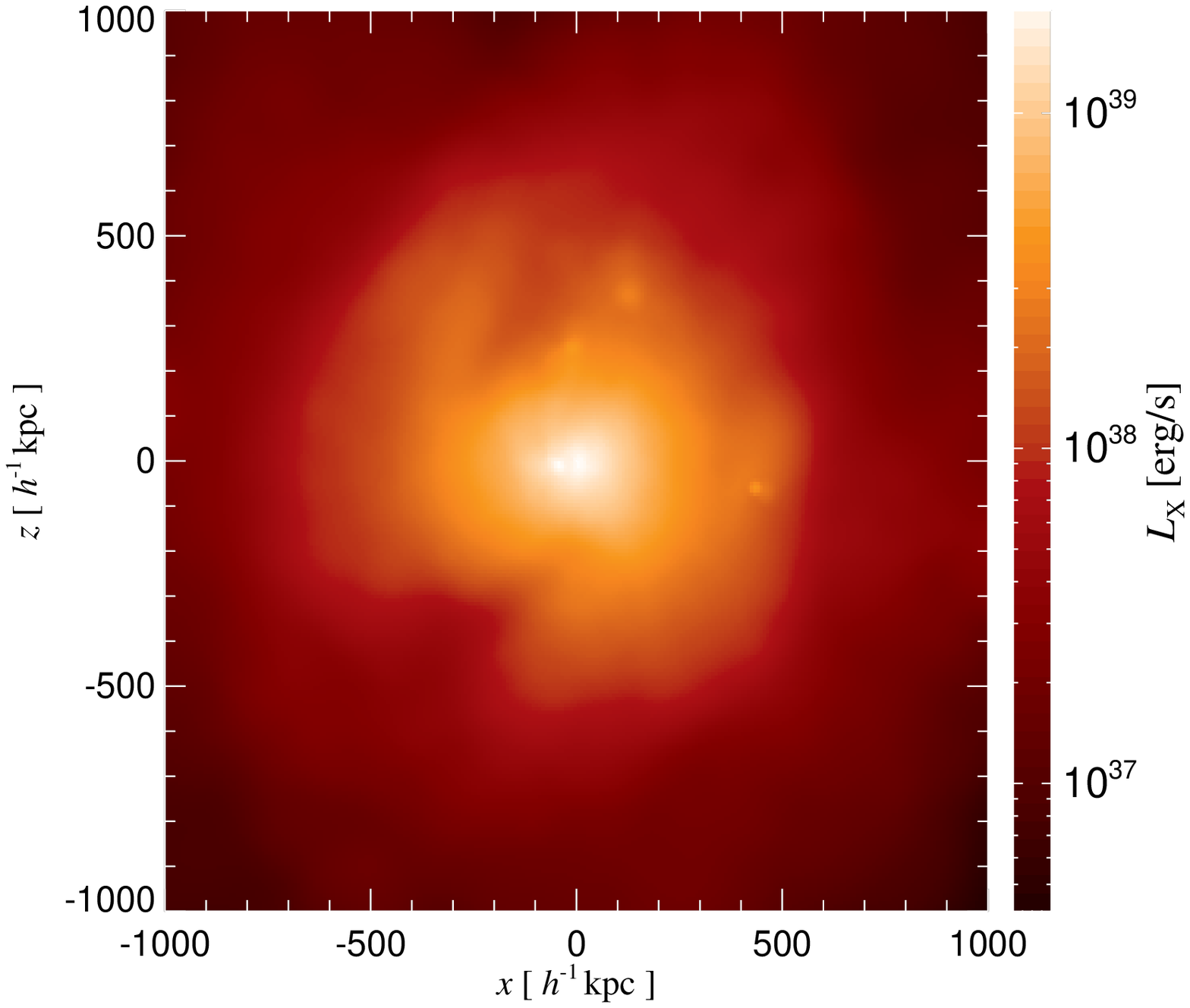,width=9.truecm,height=8.3truecm}
\hspace{-0.3truecm}
\psfig{file=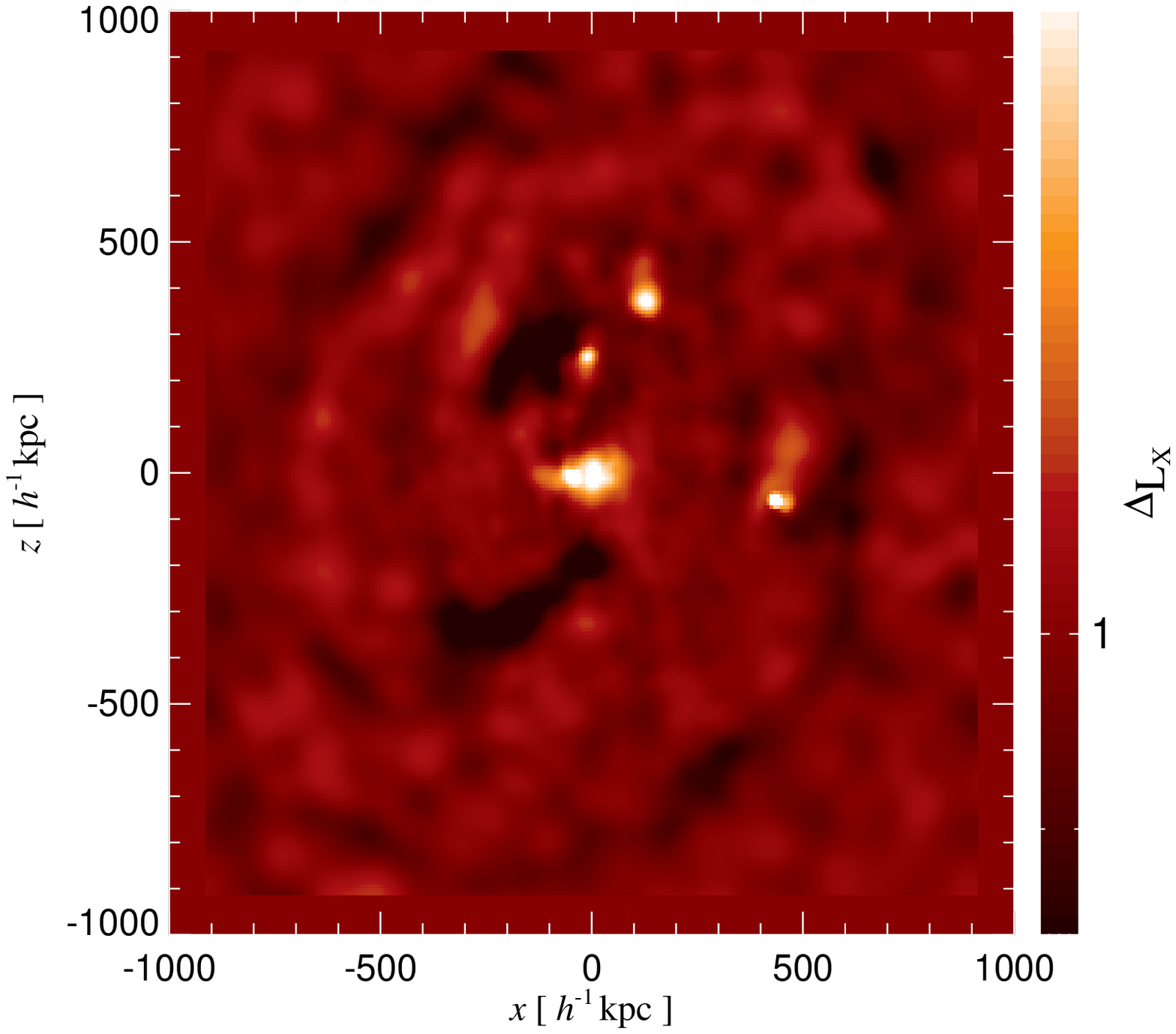,width=9.truecm,height=8.3truecm}}}}
\caption{Maps of the central region of the $g676$ galaxy cluster at
$z=0.1$. In the upper row, a mass-weighted temperature map is shown in
the left-hand panel, while the right-hand panel illustrates the
mass-weighted spatial distribution of $P_{\rm CR}$ relative to the total
pressure. In the lower row, X--ray luminosity maps are illustrated. The
right-hand panel gives an unsharp masked luminosity map, allowing for
easier bubble identification.}
\label{Lmaps}
\end{figure*} 

\subsection{Importance of Coulomb and hadronic losses}\label{Importance}

There are two primary dissipation processes for the CR particle
population. One is given by individual electron scatterings in the
Coulomb fields of CR particles and by small momentum transfers through
excitations of plasma oscillations, which is especially
efficient in the low momentum tail of the CR population
\citep{Gould1972}. The other important CR energy loss process is due to
inelastic collisions of CRs on nucleons, where hadronic interactions
produce pions that are then ultimately dissipated into radiation as well
as thermal heat \citep[e.g.][]{Ensslin2007}. Additionally, the process of
MHD wave-mediated cosmic-ray heating \citep{Loewenstein1991} could be relevant
for the CR particle population within the bubbles \citep[see e.g.][]{Guo2007},
that we do not consider in this work.

In order to gauge the relative importance of these loss processes we
have performed two additional test runs where we have switched off CR
losses altogether, or alternatively only allowed for Coulomb
cooling. Based on these runs we find that in the case where bubbles are
filled exclusively with CRs and where no diffusion is allowed, hadronic
losses are not very important during the whole simulated timespan.

Instead, we find that Coulomb cooling is especially important in the early phases
of BH growth. Here the Coulomb cooling time of the injected CR particles
turns out to be rather short such that the CR energy is quickly
thermalized into the ICM. Thus, in the run where only Coulomb cooling is
allowed the BH is accreting mass as efficiently in the beginning as in
the case with thermal feedback. However, for $t > 0.1\,t_{\rm Hubble}$,
the CR pressure inside of the bubbles builds up and the characteristic
time of Coulomb losses increases. As a result, the CRs thermalize very
little of their energy content for $t > 0.15\,t_{\rm
Hubble}$. Interestingly, in the case where no losses were allowed for the
non-thermal population BH growth is somewhat faster during the very
first part of the simulation than in the case of thermal
bubbles. Here, CRs cannot thermalize part of 
their energy into the surrounding ICM and heat it, and since the
composite of non-thermal and thermal pressure is more compressible, AGN
feedback becomes somewhat less efficient than the purely
thermal bubbles in the early stages of BH growth.

We have further explored the case where only one part of the available
mechanical energy is delivered into CRs, while the other remained in
thermal form. For values of $f_{\rm CR}$ as low as $0.1$, the influence
of CRs is barely distinguishable from the purely thermal case. However,
if we select $f_{\rm CR}$ of order of $0.5$ or higher the composite of
CRs plus thermal gas in the bubbles shows interesting properties. In
particular, for $t < 0.1\,t_{\rm Hubble}$, the non-thermal component
inside of the bubbles only weakly increases the gas compressibility, and
these composite bubbles essentially behave as pure thermal bubbles. On
the other hand, for $t > 0.1t_{\rm Hubble}$, the cosmic ray pressure
$P_{\rm CR}$ inside of the composite bubbles reaches high values, so
that CR processes become much more relevant. At this stage, the
properties of composite bubbles and their feedback effects become very
similar to the case of pure CR bubbles.

\section{Cosmic ray bubbles in simulations of cluster formation} \label{Cosmological}

In this section we consider self-consistent cosmological simulations of
galaxy cluster formation subject to AGN heating. From a cosmological box
of size $479\,h^{-1}{\rm Mpc}$ on a side that has been evolved adopting
a $\Lambda$CDM cosmology \citep{Yoshida2001, Jenkins2001}, a galaxy
cluster with a mass of $\sim 10^{14}\,h^{-1} {\rm M}_\odot$ has been
selected (the `g676' cluster), that is fairly relaxed at the present
epoch. Using the Zoomed Initial Condition (ZIC) technique
\citep{Tormen1997}, new initial conditions for this cluster have been
constructed by resampling its Lagrangian region at a higher numerical
resolution \citep{Dolag2004}. In \citet{Sijacki2007}, we have presented
several simulations of this cluster with different physics, including
runs with cooling and star formation processes only, as well as
simulations that analyzed the impact of our AGN feedback model with
thermal bubbles. Here, we discuss analogous simulations that focus on
the role of CRs in AGN-inflated bubbles. Further details about the
numerical setup can be found in \citet{Sijacki2007}. In our new
simulations, we keep all numerical parameter choices (like
gravitational softening lengths, number of SPH smoothing neighbours,
etc.) exactly the same in order to clearly pin down the impact of the
non-thermal nature of the bubbles.

\subsection{Signatures of CR bubbles}\label{Observational}

We start our analysis by trying to identify a link between the CR bubble
properties and specific features they induce in the ICM. In
Figure~\ref{Lmaps}, we show the central region of our most massive
cluster at $z=0.1$. The thickness of the projection slice is $300
\,h^{-1}{\rm kpc}$, and the different panels give the projected
mass-weighted temperature map (upper left-hand panel), the $P_{\rm
CR}$-to-total pressure ratio (upper right-hand panel), the X--ray
luminosity map (lower left-hand panel) and an unsharp masked luminosity
map (lower right-hand panel). The unsharp masked luminosity map has
been produced by dividing the original luminosity map by a version of
it that has been smoothed on $\sim 180 \,h^{-1}{\rm kpc}$ scale. In
the upper right-hand panel, the 
position of CR bubbles can be clearly identified. In the very centre of
the cluster there is significant cosmic ray pressure, corresponding to
very recent BH activity which is producing two small bubbles. Moreover,
upwards from the central region there is a large plume filled with CR
protons extending from $\sim 200 \,h^{-1}{\rm kpc}$ to $\sim 500
\,h^{-1}{\rm kpc}$ along the $z$-axis. Yet another plume is clearly
visible in the lower region, propagating towards the lower-left corner,
and having a total length of more than $\sim 200 \,h^{-1}{\rm kpc}$. The
temperature map reveals asymmetries along the direction of big CR
plumes, and a close correspondence to the central small bubbles, where
the temperature is somewhat reduced. However, it is in the luminosity
maps that one can clearly identify CR bubbles, as depression regions in
X--ray emissivity. The bubble morphologies and the reduction in X--ray
emissivity in the regions that are permeated with CR particles are in
very good qualitative agreement with a number of observational findings
\citep[e.g][]{Owen2000, Nulsen2005, Forman2006, Sanders2007}, suggesting
that the presence of a non-thermal component inside of the bubbles is
essential for more faithful reproduction of the observational results
and their understanding.

\begin{figure*} \centerline{ \vbox{\hbox{
\psfig{file=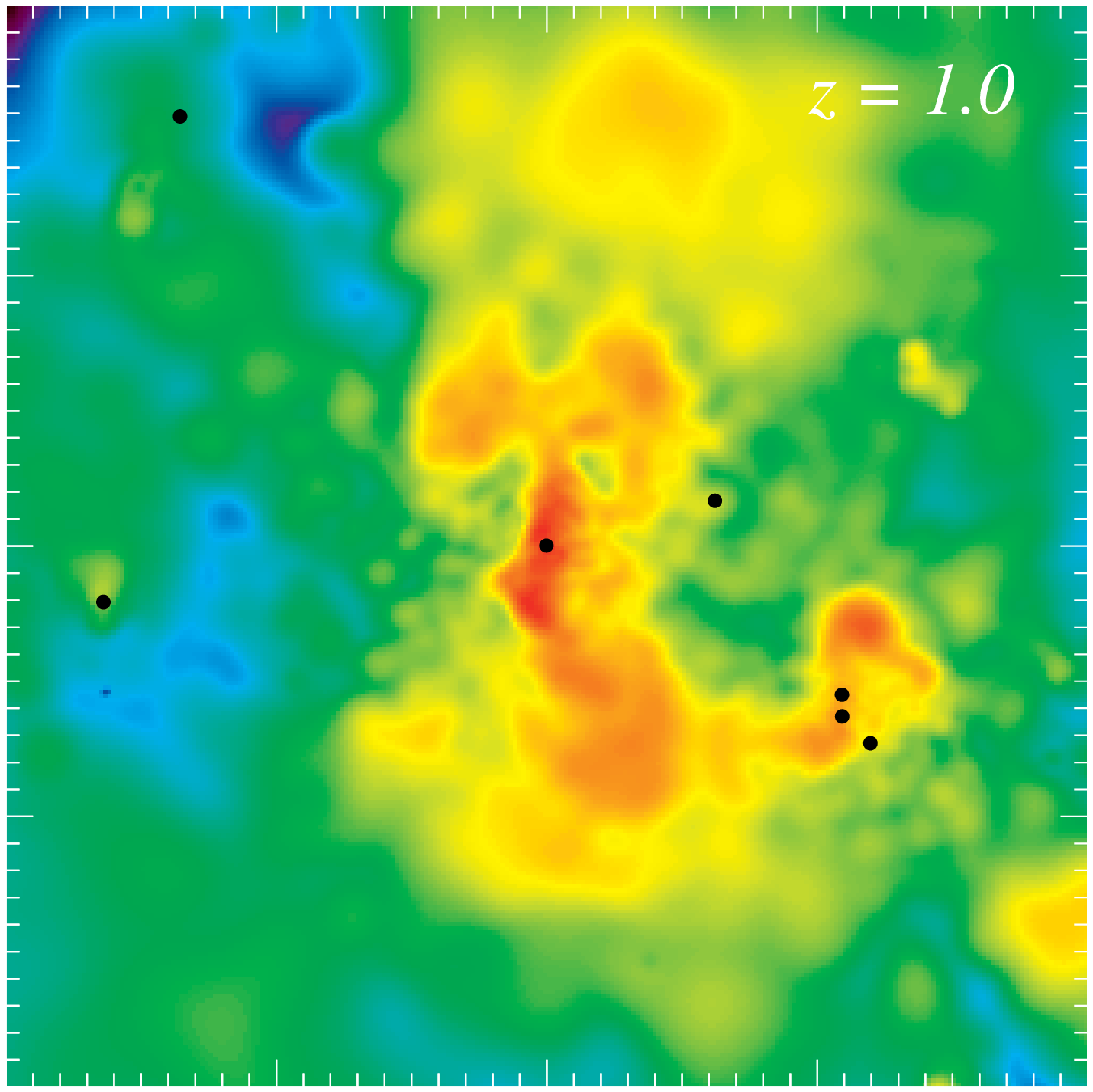,width=7.5truecm,height=7.truecm}
\hspace{-0.5truecm}
\psfig{file=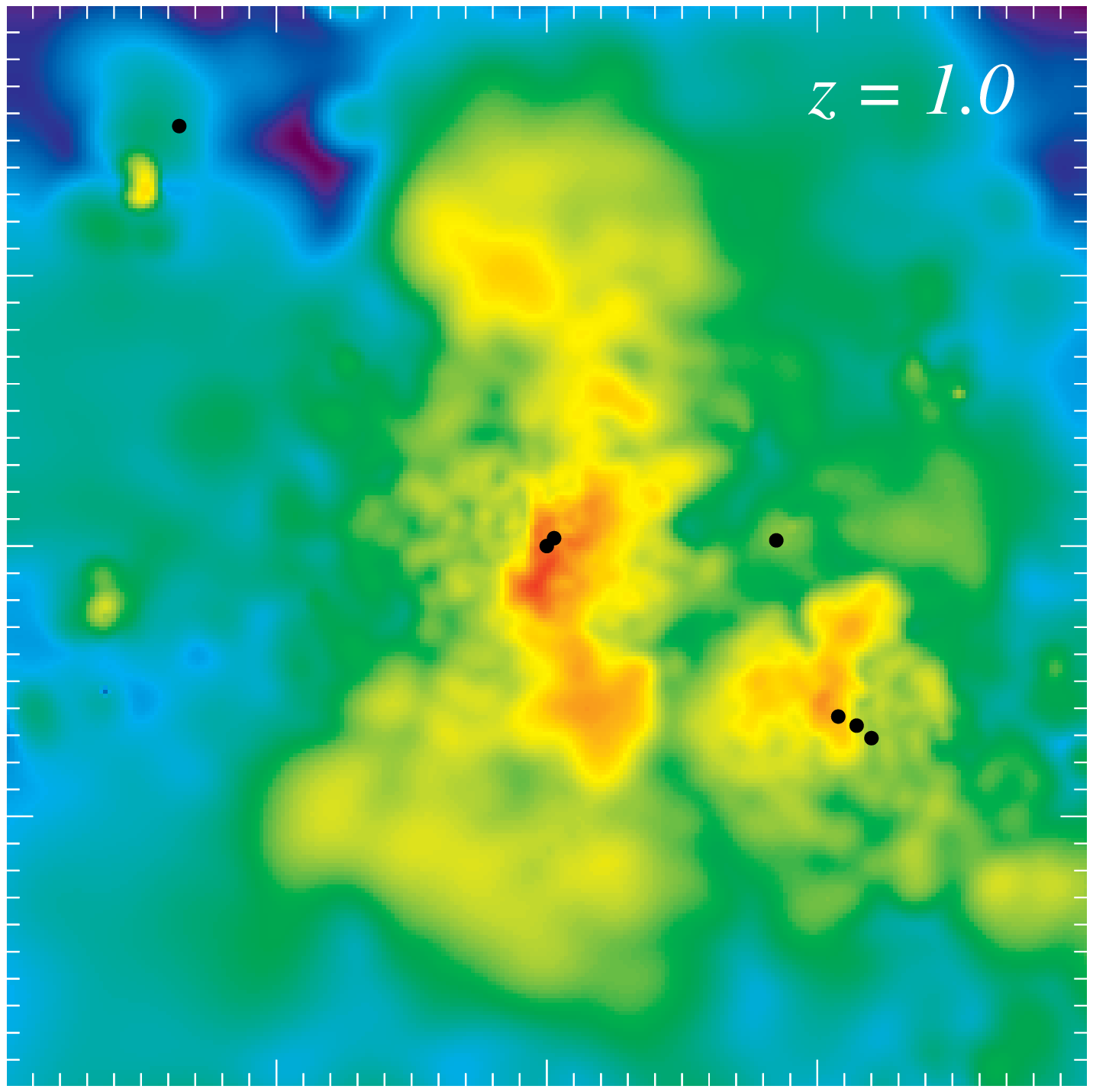,width=7.5truecm,height=7.truecm}
\hspace{-0.5truecm}
\psfig{file=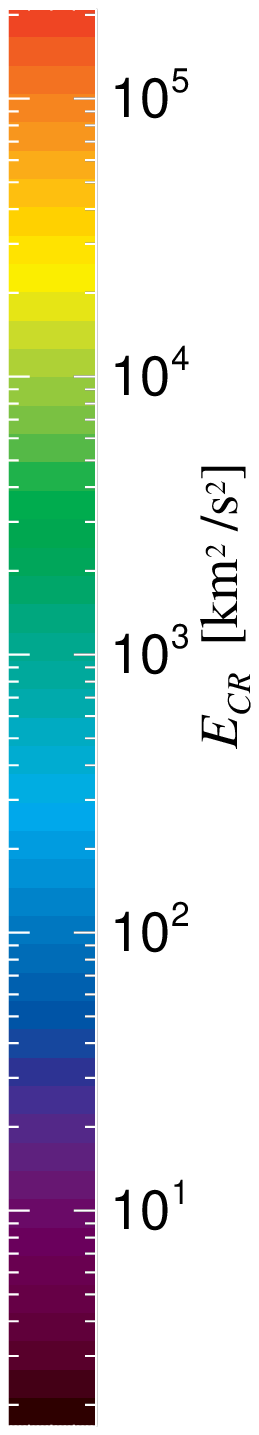,width=1.3truecm,height=7.truecm}}
\vspace{-0.3truecm} \hbox{
\psfig{file=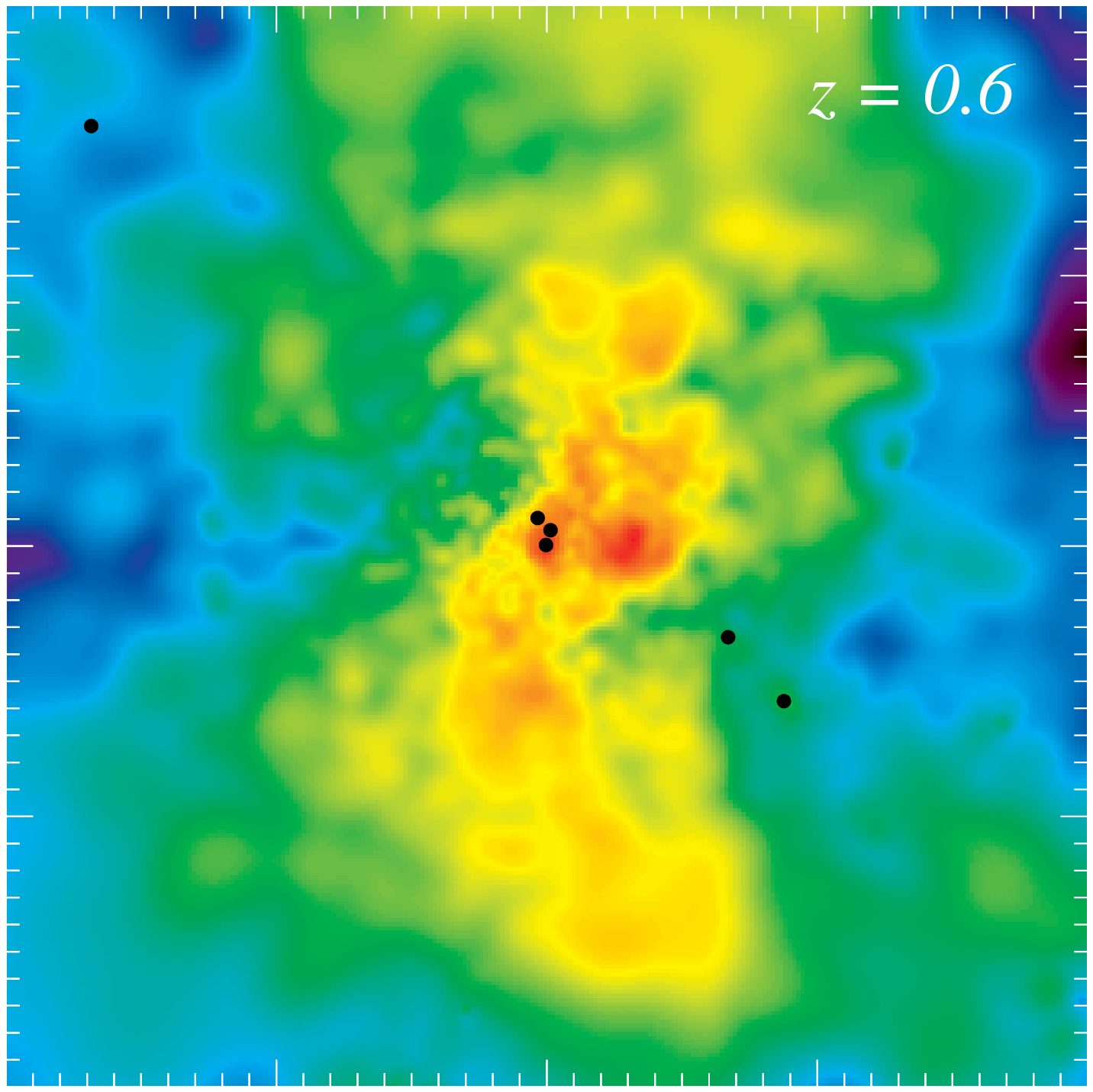,width=7.5truecm,height=7.truecm}
\hspace{-0.5truecm}
\psfig{file=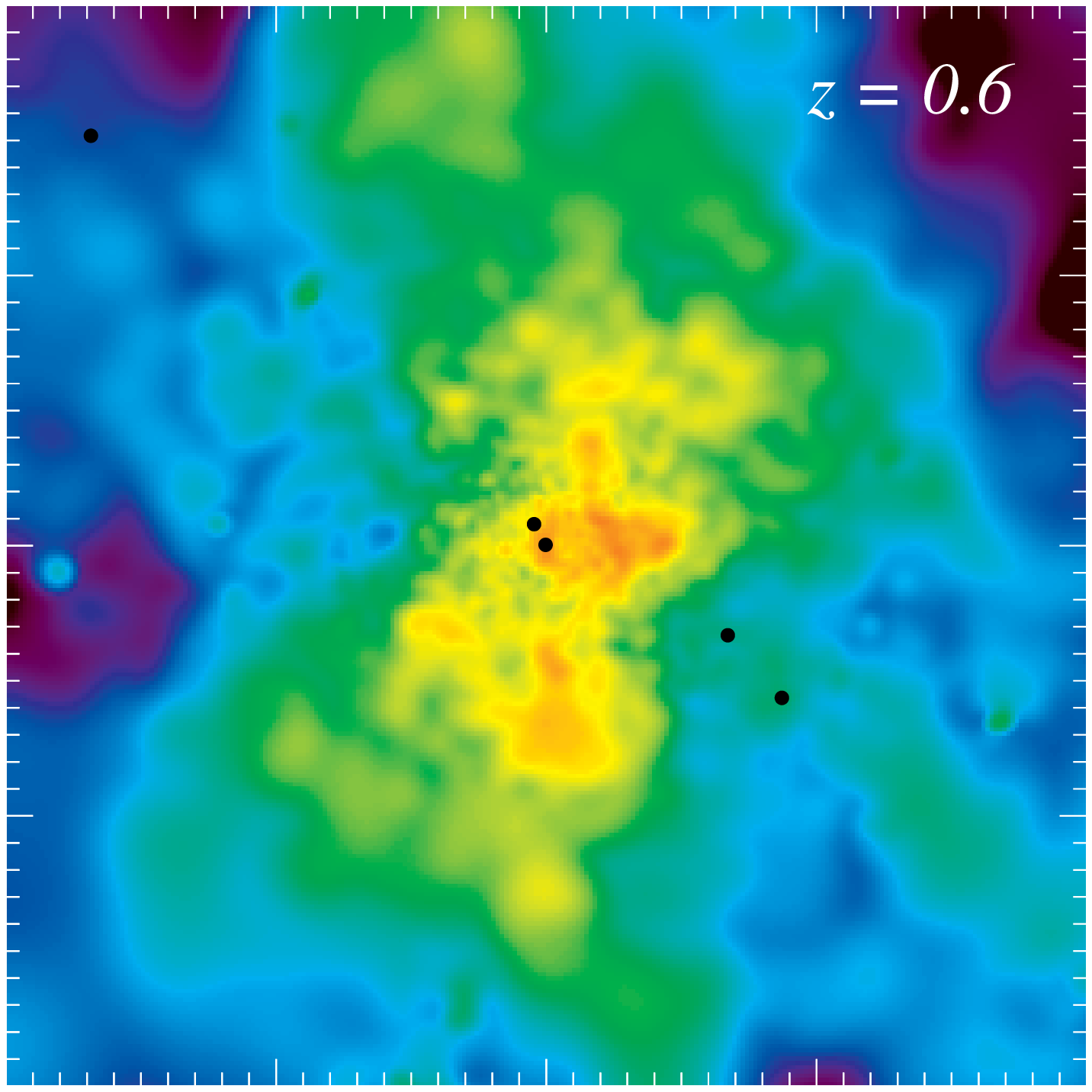,width=7.5truecm,height=7.truecm}
\hspace{-0.5truecm}
\psfig{file=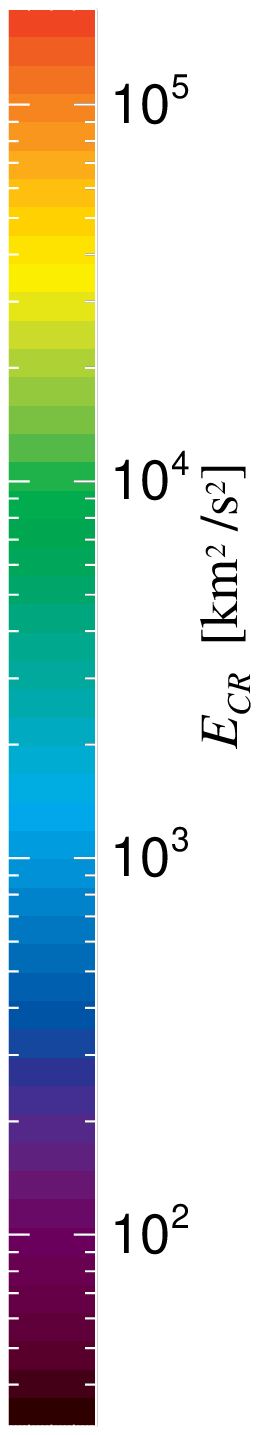,width=1.3truecm,height=7.truecm}}
\vspace{-0.3truecm} \hbox{
\psfig{file=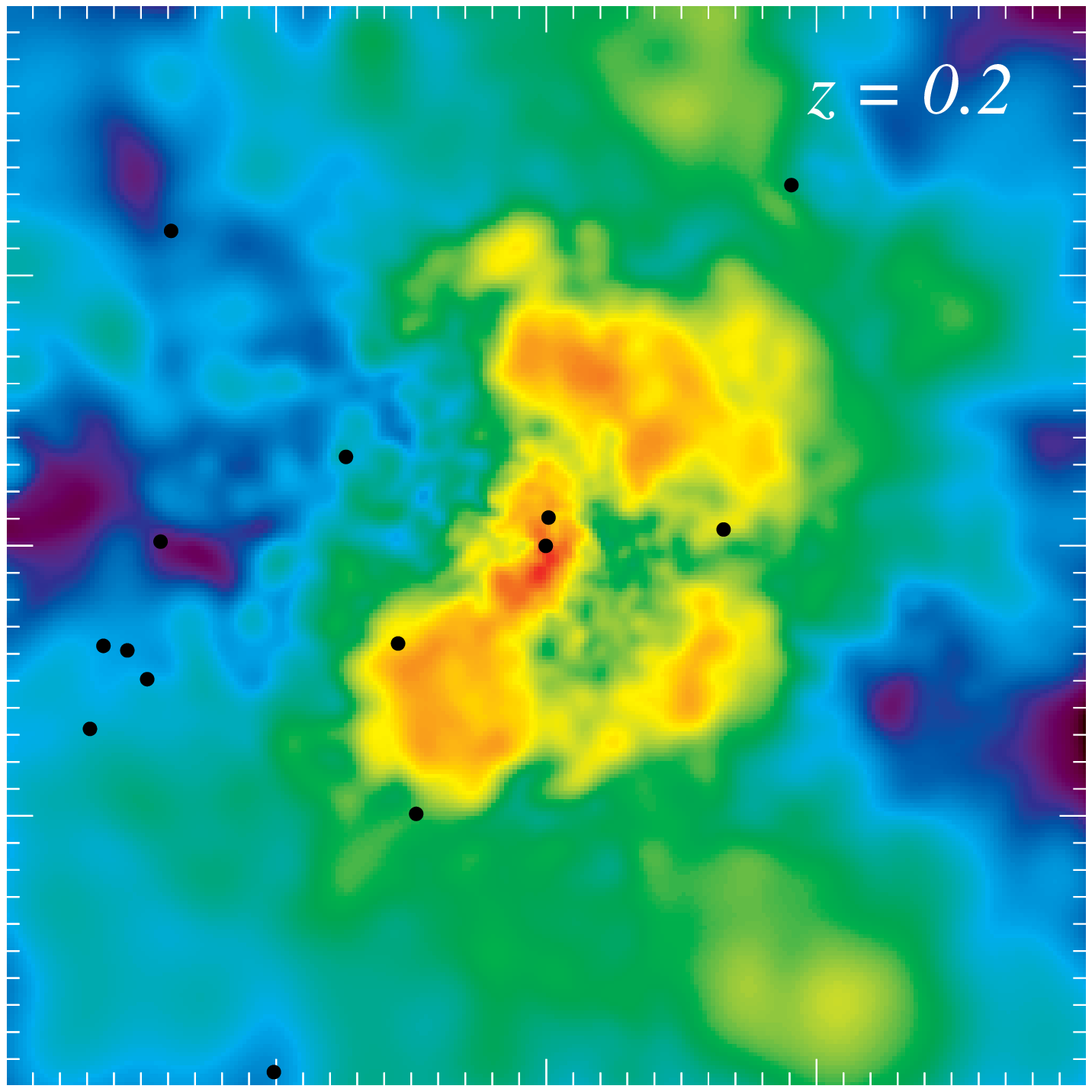,width=7.5truecm,height=7.truecm}
\hspace{-0.5truecm}
\psfig{file=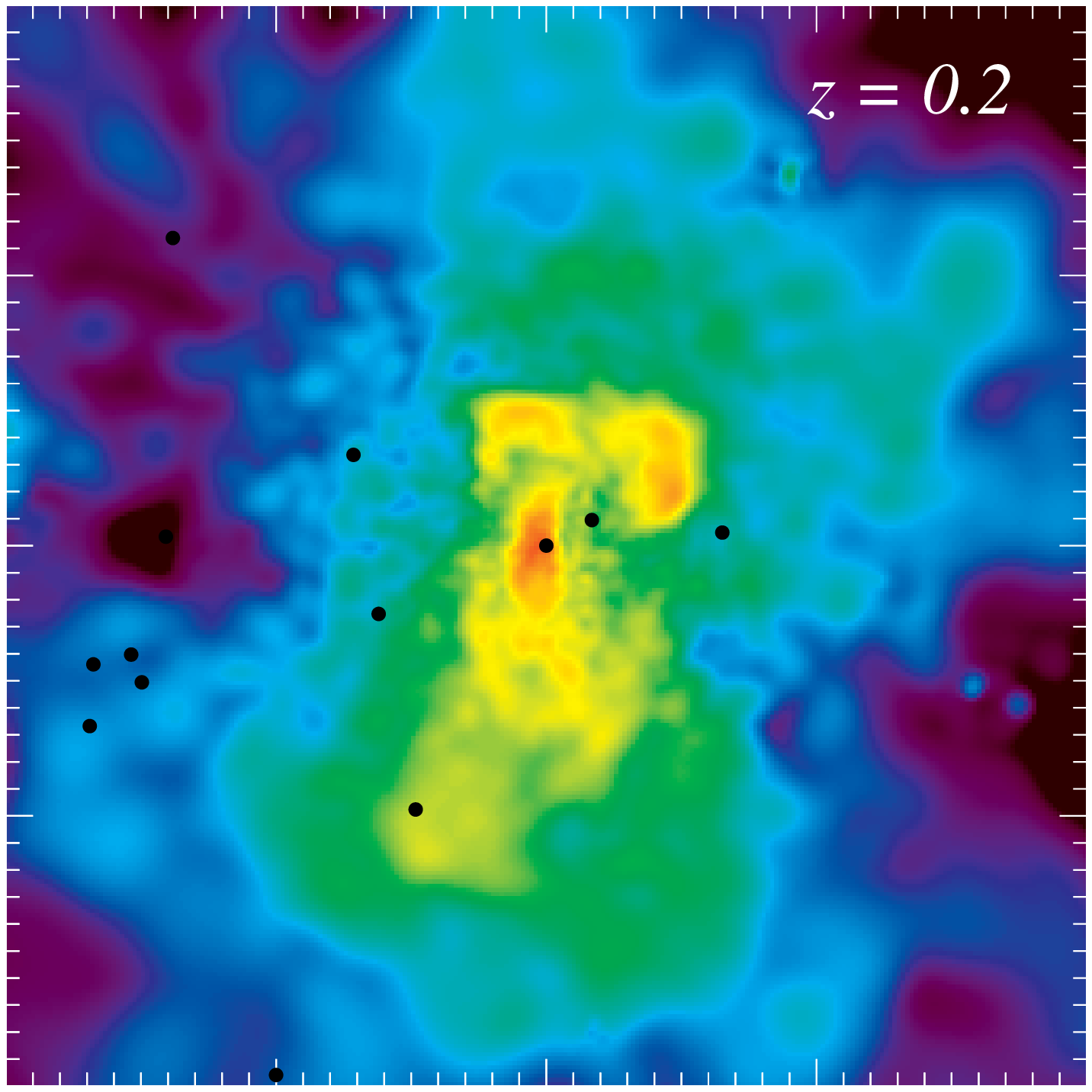,width=7.5truecm,height=7.truecm}
\hspace{-0.5truecm}
\psfig{file=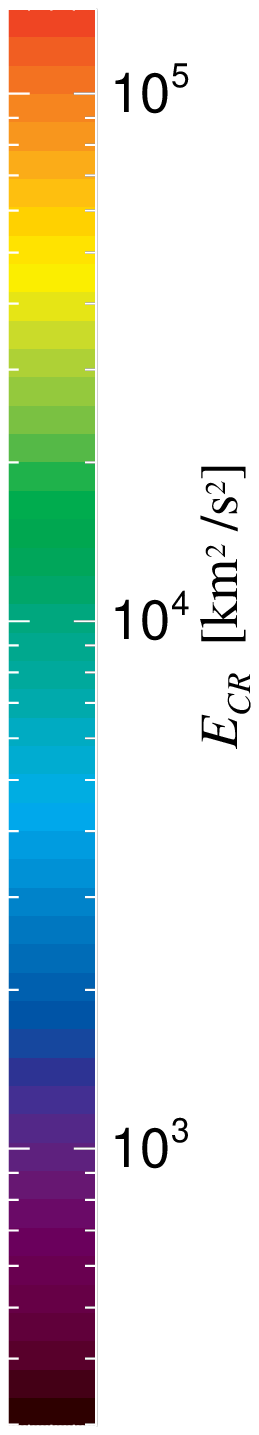,width=1.3truecm,height=7.truecm}}}}
\caption{Projected mass-weighted CR energy per unit mass in boxes of $4000
\,h^{-1}{\rm kpc}$ on a side at three different redshifts: $z=1.0$
(top), $0.6$ (middle) and $0.2$ (bottom). The left-hand panels are for
the run with $\alpha=2.1$, while the right-hand panels show the case of
$\alpha=2.4$. The selected redshifts correspond to highly perturbed
states of the cluster when it is undergoing merging events. It can be
seen how environmental effects influence the morphologies and dynamics
of CR bubbles.}
\label{Emaps}
\end{figure*}

In Figure~\ref{Emaps}, we analyze the CR bubble morphologies at
different cosmic times. To this end we plot the projected CR energy
per unit mass in
the $x$-$z$ plane within a cubic box of size $4000 \,h^{-1}{\rm kpc}$ on
a side, at three different redshifts, $z=1.0$, $0.6$ and $0.2$. The
left-hand panels refer to the run performed with $\alpha=2.1$, while the
right-hand panels are for $\alpha=2.4$. The black dots over-plotted on
the maps mark the positions of BH particles more massive than $1.5\times
10^7 \,h^{-1} {\rm M}_\odot$. Note that we have deliberately chosen not
to vary the injection axis of the bubbles with time (which is always
along the $z$-axis). This has been done in order to assess to what
extent environmental effects like bulk motions, local overdensity
fluctuations, or merger events, can displace bubbles from their initial
injection position. At the first glance it can be noticed that the
evolved CR bubble morphologies are rather complex, clearly lacking a
strong axial symmetry, even though they were injected in a symmetric
fashion. This emphasizes the important role played by the cosmological
environment in modifying the bubble shapes. 

The three epochs shown in Fig.~\ref{Emaps} have been selected such that
they correspond to merging events of the central BH with a smaller mass
BH that once belonged to a satellite halo. Thus, at these epochs, the BH
activity in the central region is rather vigorous, and several bubble
episodes can be seen, from the central ones that are just injected, to
the more peripheral ones which have risen and are now detached, but
still have significant pressure support from CR protons. Interestingly,
at $z=1.0$, in the vicinity of the most massive cluster in the
lower-right part, there is a smaller halo containing a central massive
BH which during a similar time span has also generated two bubbles. At
$z=0.6$ this smaller halo reaches $\sim 1/3$ of the central cluster mass
and merges with it violently. The central BHs of these halos merge as
well, increasing the mass of the most massive BH by $\sim 10^9 \,h^{-1}
{\rm M}_\odot$. It can be seen that during this merger event the relic
CR bubbles are significantly perturbed and displaced towards the right,
in the direction from which the smaller halo was falling in. 

Finally, by
comparing the left-hand panels to the right-hand panels in
Fig.~\ref{Emaps}, it can be seen that CR bubbles with a steeper
power-law spectrum have in general similar morphologies to the
$\alpha=2.1$ case, but they show a somewhat smaller size and shorter
survival time, as expected given that Coulomb losses become more
important for a steeper spectrum.

\subsection{Impact of CR bubbles on their host haloes}

\subsubsection{Radial profiles}

\begin{figure*} \centerline{
\psfig{file=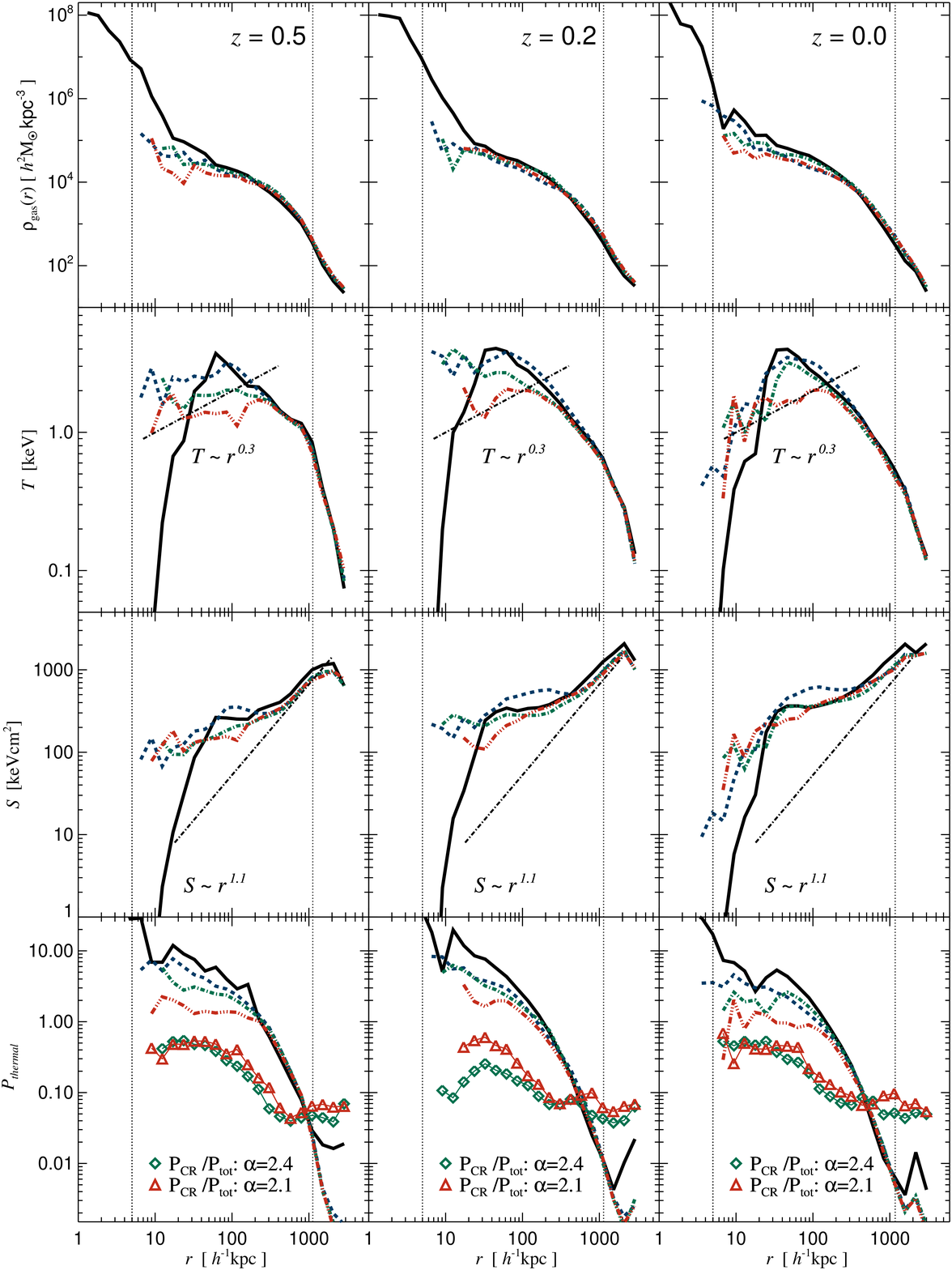,width=17truecm,height=21truecm}
}
\caption{Radial profiles of gas density (first row), mass-weighted
temperature (second row), entropy (third row) and thermal pressure (last
row, in internal code units) of the g676 galaxy cluster at $z = 0.5$, $0.2$ and $0.0$,
respectively. Solid black lines illustrate the case without AGN
heating. The dashed blue lines are for simulations where AGN feedback is
included with purely thermal bubbles, as in \cite{Sijacki2007}. The runs
with CR filled bubbles are represented with green dot-dashed lines and
red triple dot-dashed lines, for spectral indices of $\alpha=2.4$ and
$\alpha=2.1$, respectively. The vertical dotted lines denote the
gravitational softening length and the virial radius of this cluster.
The dash-dotted lines in the second row of panels show the slope of the
central temperature profiles of the cool core clusters found by
\cite{Sanderson2006}, while the dash-dotted lines in the third row
illustrate the entropy scaling with radius, i.e. $S \propto r^{1.1}$. In
the last row we also indicate the mean ratio of CR-to-total pressure in
each bin, which is shown with green diamonds ($\alpha=2.4$) and red
triangles ($\alpha=2.1$).}
\label{g676_profiles}
\end{figure*} 

In Figure~\ref{g676_profiles}, we show radial profiles of gas density
(first row), mass-weighted temperature (second row), entropy (third
row), and thermal pressure (last row) at three different epochs, $z=0.5$,
$0.2$ and $0.0$. The black continuous lines are for the run with cooling
and star formation only, blue dashed lines denote the run with thermal
bubbles \citep[as in][]{Sijacki2007}, while the green dot-dashed
($\alpha=2.4$) and red triple dot-dashed lines ($\alpha=2.1$) show the
case with CR bubbles. It can be seen that the gas density is affected by the
presence of relativistic particles in the 
bubbles, being roughly comparable to the simulation where the bubbles are purely
thermal. However, the gas temperature is altered substantially and in
general is always reduced in central regions, as we have also found for
our isolated cluster simulations. This is simply caused by the fact that
significant pressure support in the central cluster regions is in a
non-thermal form, which decreases the thermal pressure needed to
counteract the gravity exerted by the cluster's potential. In
\citet{Sijacki2007} we have already shown that a self-regulated AGN
feedback mechanism is required in order to obtain realistic temperature
profiles that can reproduce the decline towards the innermost regions
that has been observed in a number of cool core clusters. Having a
significant population of non-thermal particles in the central bubbles
further helps in this respect, and provides a natural explanation for
the observed shape of the temperature profiles. Thus, the presence of a
relativistic particle population inside AGN-inflated bubbles is not only
essential for reproducing the observed morphologies of bubbles and X-ray
cavities in the hot ICM, but it also improves the temperature
distribution of the simulated galaxy clusters. However, our results also
show that the AGN feedback efficiency and the assumed fraction of the
energy in high energy CR protons are to a certain extent degenerate,
suggesting that a detailed comparison with observations is needed to
constrain both of these parameters individually.

The CR bubble feedback also steepens the entropy profile and brings it into
closer agreement with X-ray observations, especially at intermediate radii
from $\sim 20 \,h^{-1}{\rm kpc}$ to $\sim 300 \,h^{-1}{\rm kpc}$. Further
information about the role of CR bubbles can be obtained from the last row of
panels in Figure~\ref{g676_profiles}, where we plot radial profiles of the
mass-weighted thermal pressure for all the runs we performed. The symbols give
the mean ratio of CR-to-total pressure in each spherical bin. It can be seen
that the CR pressure is most relevant in the central regions, being comparable
to the thermal pressure for $r < 50 h^{-1}{\rm kpc}$ at all three epochs
considered. At $z=0$ and inside $100 h^{-1}{\rm kpc}$, the mean CR pressure
contributes about $37\%$ of the total pressure for a spectral slope of
$\alpha=2.1$ (for the case of $\alpha=2.4$ this number is $\sim 23\%$), while
at the virial radius, the non-thermal pressure support drops to $14\%$ (or
$10\%$ for $\alpha=2.4$). Recall however that in this study we have excluded
other potential sources of CRs, namely supernovae and cosmic structure
formation shocks. They are expected to further increase the overall CR
pressure content, and may also affect the relative contributions of
non-thermal to thermal pressure as a function of radius \citep[for further
details on these issues see][]{Pfrommer2007a}. Interestingly, given that
the CR pressure contribution is comparable to the thermal pressure and that
the CR pressure gradient is negative, the central cluster gas might be
convectively unstable as proposed by \citet{Chandran2006} and explored in the
case of SNe injected CRs by \citet{Pfrommer2007a}. Actually, since most of the
CR pressure is due to CRs residing in the buoyantly rising bubbles, one can
argue that the system indeed exhibits a CR-driven convection. An instability
did not need to develop, since the CRs in the ICM are directly injected into
an unstable configuration of a bubble.

The total X-ray luminosity within the virial radius in the case of CR
bubbles with $\alpha=2.4$ lies in between the simulations performed
without AGN feedback and the run with thermal bubbles. This increase of
the X-ray emission is a result of the substantial Coulomb losses of the
CR population combined with the softer equation state of the gas
supported by non-thermal pressure. In the run with $\alpha=2.1$,
Coulomb losses are less important and the total X-ray luminosity is
reduced from a value of $\sim 1.6\times10^{43}\,{\rm erg/s}$ for the
case without AGN feedback to $\sim 9.2\times10^{42}\,{\rm erg/s}$. Also,
in the simulation with $\alpha=2.1$, the mean mass-weighted gas
temperature is reduced, from $1.20\,{\rm keV}$ to $0.97\,{\rm keV}$,
indicating that CR bubbles may significantly influence galaxy cluster
scaling relations such as the $L_{\rm X}-T$ relationship, especially at
the low mass end. We plan to analyze this question in more detail in a
forthcoming study. Finally, we note that the bulk of the BH growth is
not affected much by the presence of a non-thermal component in the
bubbles: the BH growth is moderately reduced for $z < 2$, so that
the most massive BH in the simulated volume has its mass reduced by less
than $20\%$ at $z=0$. We also note that in our model the central star
formation rate and the total stellar mass within $R_{\rm vir}$ (see
below) are not very sensitive to whether the bubbles are filled with
thermal or non-thermal energy, indicating that feedback by CR bubbles is
about equally efficient in suppressing star formation in central
galaxies as thermal ones.

\subsubsection{Baryon fraction and AGN-driven bubbles}

\begin{figure*} \centerline{\vbox{
\psfig{file=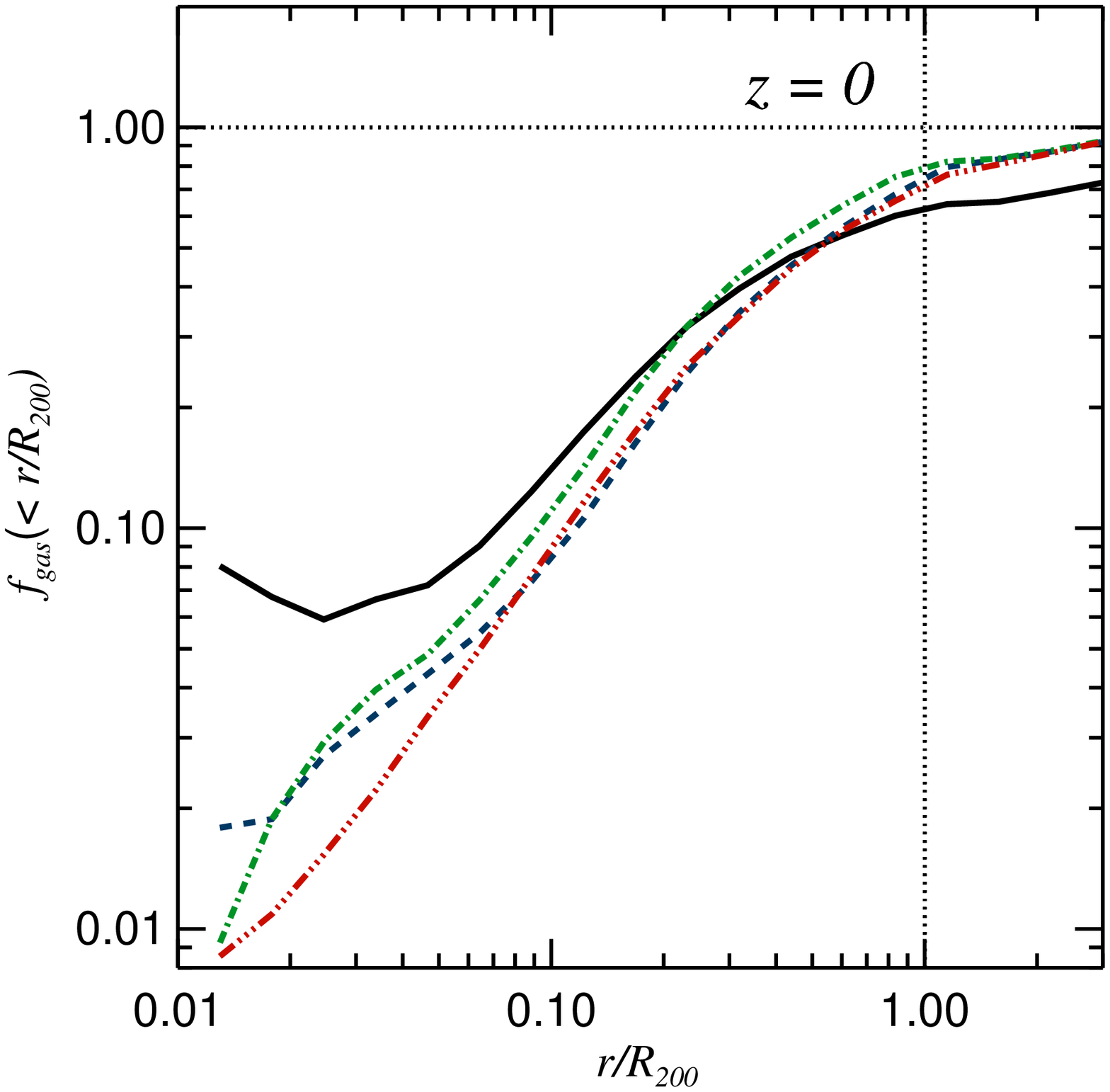,width=8.5truecm,height=8.5truecm}
\psfig{file=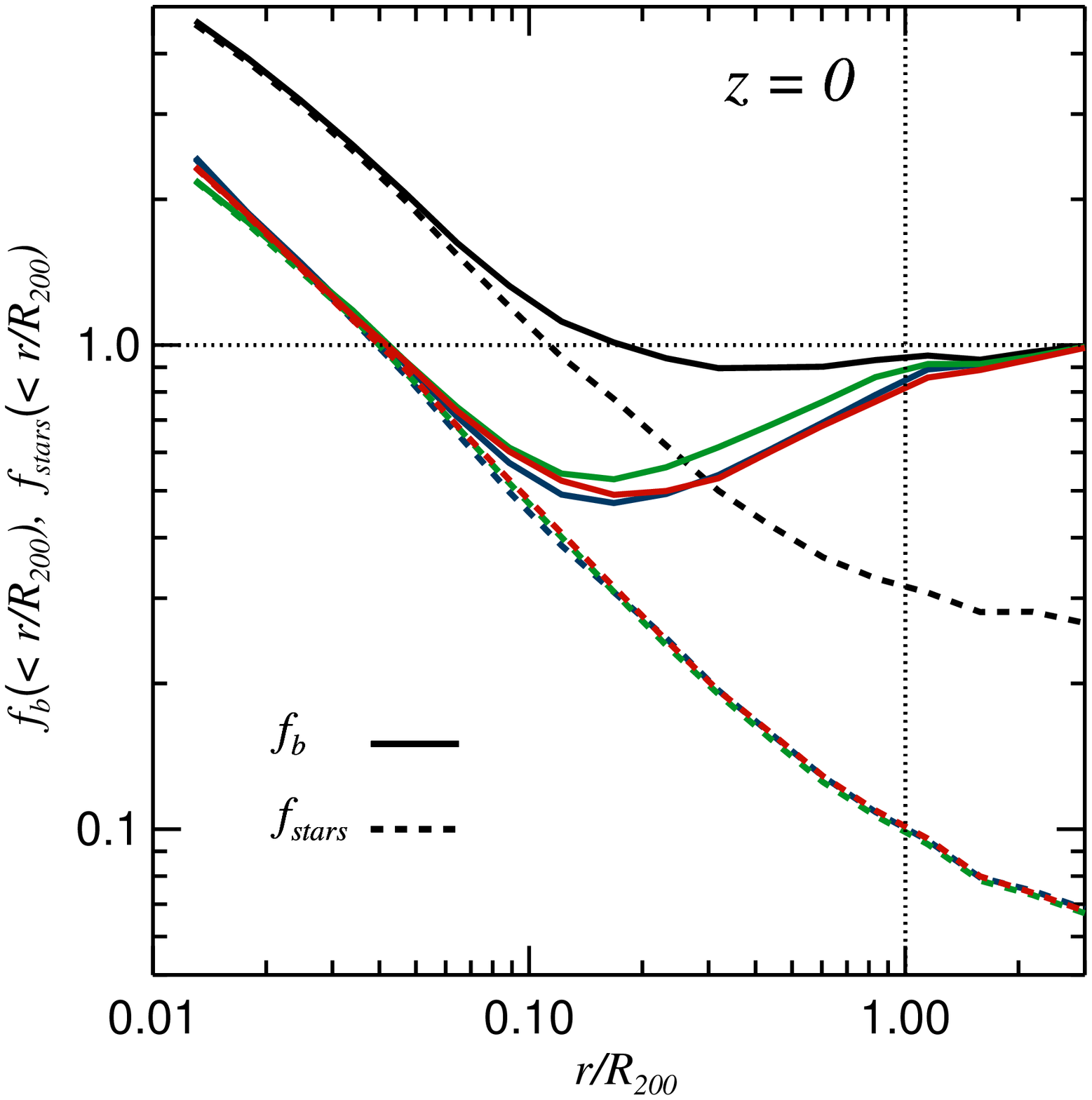,width=8.5truecm,height=8.5truecm}}}
\caption{The left-hand panel shows the profile of the cumulative gas
mass fraction as a function of radius at $z=0$. The baryon fractions
have been normalized to the universal value assumed in our simulations,
i.e. $0.13$. The black solid line is for the simulation with cooling and
star formation only, while the remaining lines illustrate results for
different runs with AGN feedback: thermal bubbles (blue dashed line), or
bubbles filled with CRs (green dot-dashed for $\alpha=2.4$, and red
triple dot-dashed for $\alpha=2.1$). The right-hand panel shows the
total baryon mass fraction (solid lines) and the stellar mass fraction
(dashed lines) with the same colour-coding.}
\label{g676_fb}
\end{figure*}

In the left-hand panel of Figure~\ref{g676_fb}, we show the gas
fraction at $z=0$ within a given radius as a function of distance from
the cluster centre. The radius has been normalized to $R_{\rm 200}$ and
the gas fraction is in units of the universal baryon fraction adopted in
our simulation (i.e. $\Omega_b/\Omega_0=0.13$). The black line is for
the run without AGN feedback, the blue line is for thermal bubbles,
while the green ($\alpha=2.4$) and red ($\alpha=2.1$) lines show the
case of CR bubbles. As a generic feature of AGN feedback, it can be seen
that the gas fraction is significantly reduced in the central cluster
region, while it is increased in the cluster outskirts. As expected, the
central gas fraction is most suppressed in the simulation where the CR
bubbles have a shallow momentum spectrum. On the other hand, the case
with $\alpha=2.4$ leads to a somewhat higher gas fraction with respect
to the simulation with purely thermal bubbles, due to the combined
effects of stronger Coulomb cooling and higher gas compressibility.

In the right-hand panel of Figure~\ref{g676_fb}, we show the total
baryon fraction (solid lines), and the stellar fraction (dashed lines)
as a function of radius, using the same colour-coding. The amount of
stars that form in the central regions is significantly reduced by the
presence of a supermassive BH, but it is not affected by the nature of
the bubbles, as discussed above. The total amount of stars drops from a
high value in excess of $\sim 30\%$ reached in the simulation without
AGN heating to a much more realistic value of $\sim 10\%$, which is
consistent with observational findings \citep{Lin2003}. This also endows
the central galaxy with a much redder colour, as observed. Due to the
reduction of the central stellar fraction the total baryon fraction is
lowered in the central regions as well. Only at radii comparable to the
turnaround radius, at $\sim 3 \times R_{\rm 200}$, it reaches the
universal value in all runs performed.

\subsubsection{Thermal SZ effect}

\begin{figure*} \centerline{\vbox{
\psfig{file=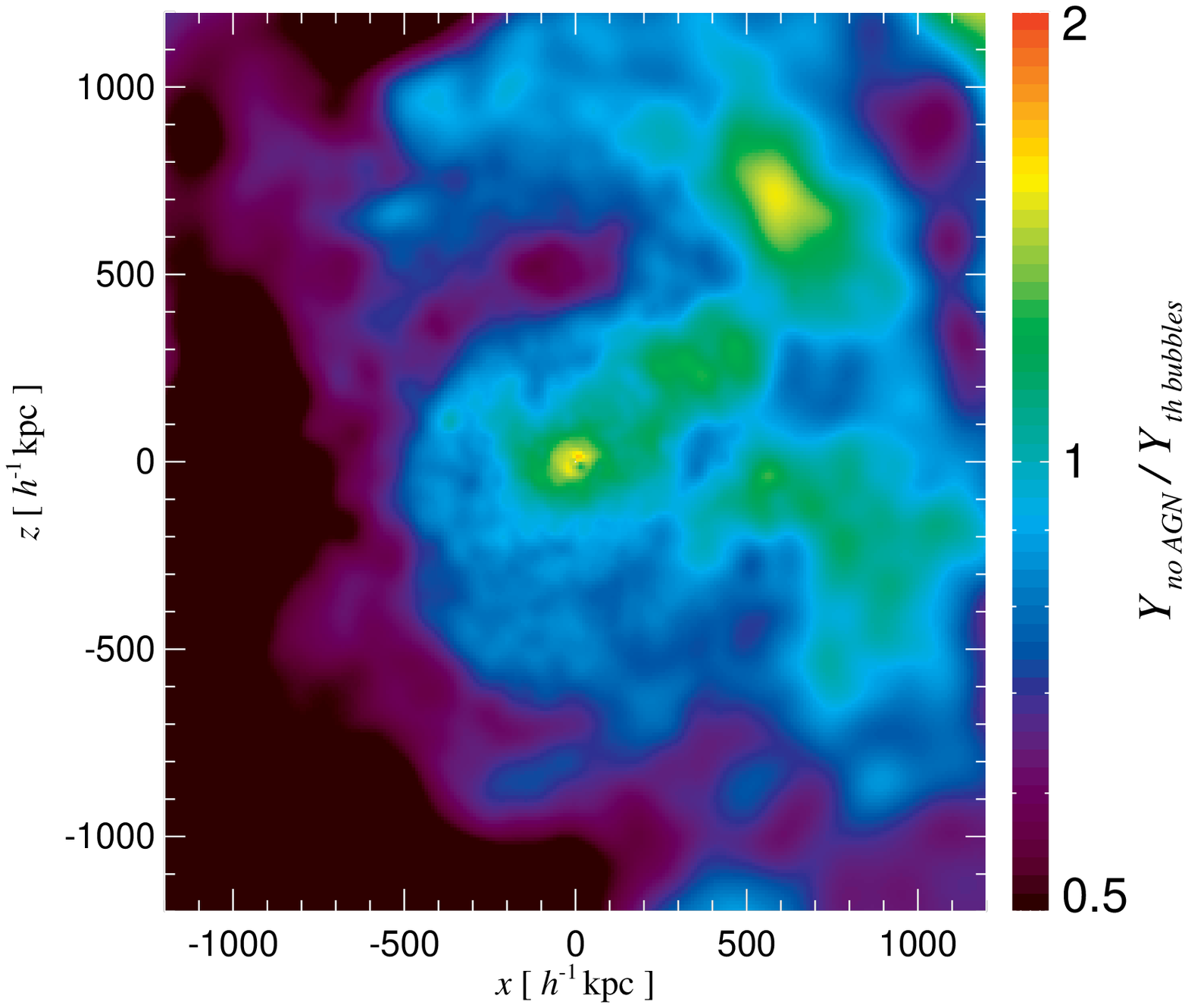,width=6.truecm,height=5.5truecm}
\hspace{-0.5truecm}
\psfig{file=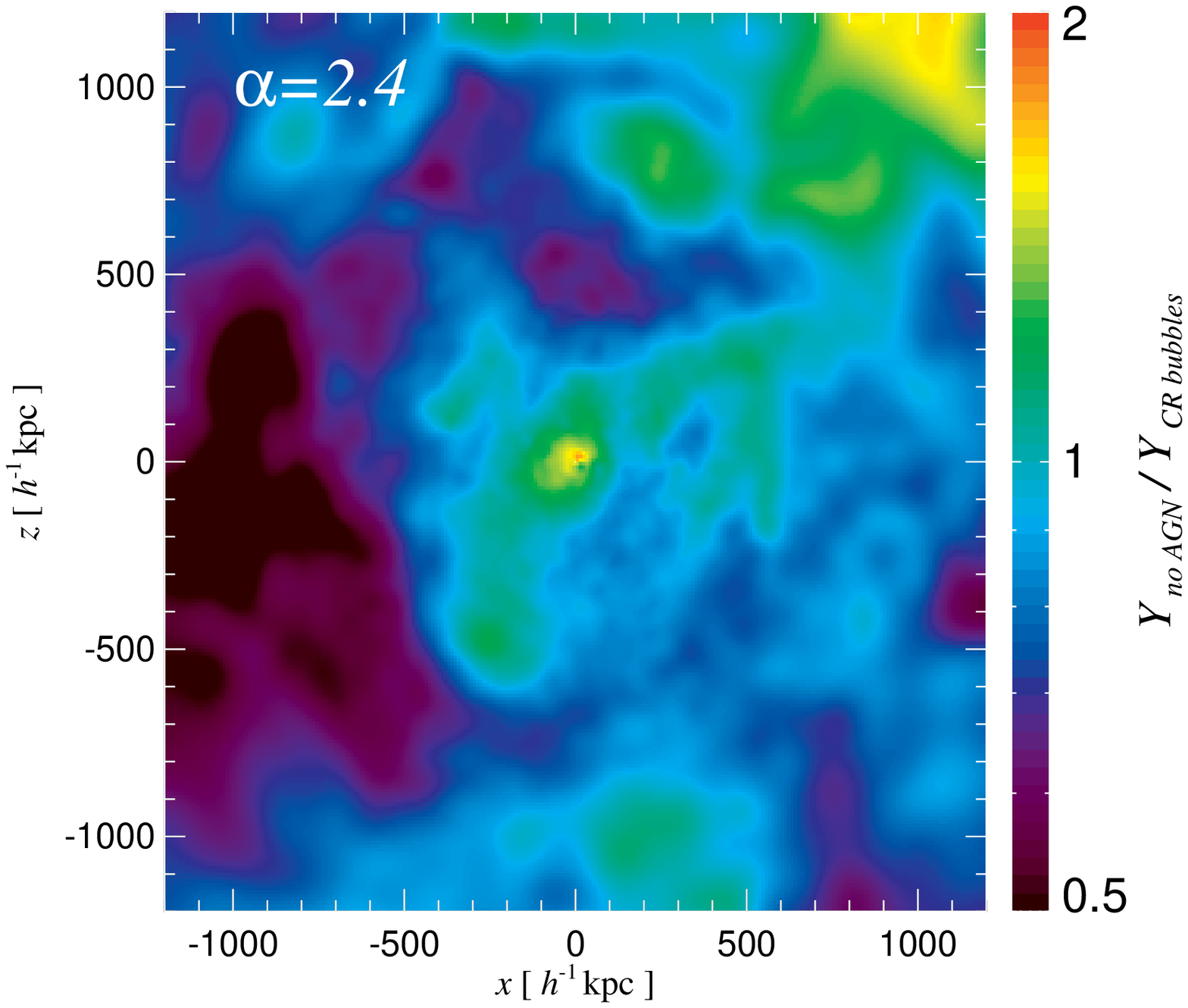,width=6.truecm,height=5.5truecm}
\hspace{-0.5truecm}
\psfig{file=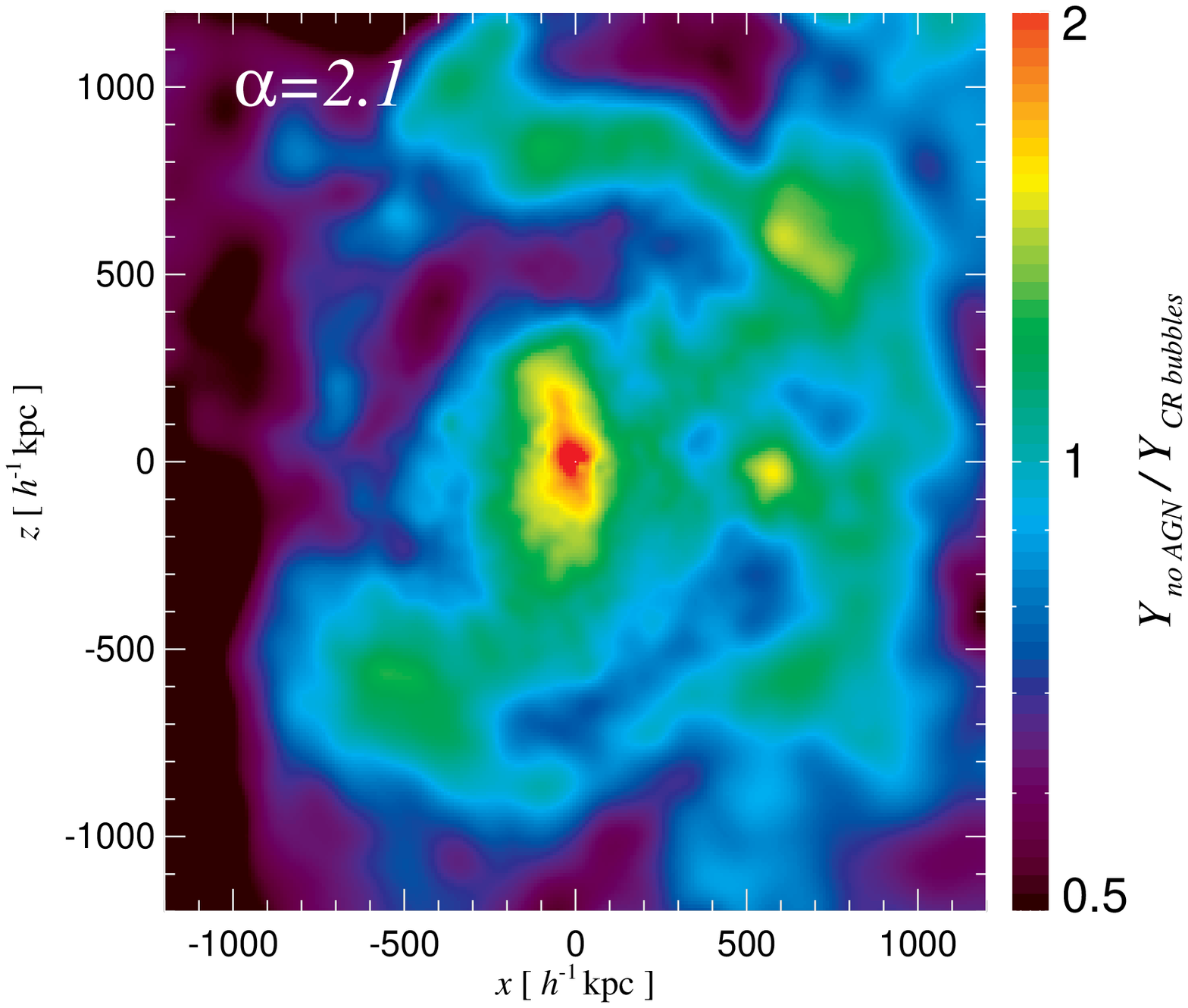,width=6.truecm,height=5.5truecm}}}
\caption{ Ratio of the Compton-$y$ parameters in our simulations without
AGN feedback and in runs with AGN-driven bubbles at $z=0$. The
left-hand panel shows the case of thermal bubbles, the central panel
that of CR bubbles with $\alpha=2.4$, and the right-hand panel the case
of CR bubbles with a steeper spectral index of $\alpha=2.1$. It can be
seen that the Compton-$y$ parameter is decreased in the central regions
due to the bubbles in all runs considered. However, in the cluster
outskirts the bubbles tend to increase the Compton-$y$ parameter, such
that the area-integrated Comptonization can be both lower or higher than
in the runs without AGN feedback.}
\label{g676_Y}
\end{figure*}

Given that CRs provide a significant pressure support in the central
cluster regions it is interesting to examine how they affect the
amplitude of the thermal Sunyaev-Zel'dovich effect\footnote{Note that we do not
  need to consider relativistic corrections to the pure thermal
  Sunyaev-Zel'dovich effect, given that we have modelled only CR protons. For
  a comprehensive study of non-thermal bubble imprints on the Sunyaev-Zel'dovich
  effect see e.g. \citet{Pfrommer2005}.}. In
Figure~\ref{g676_Y}, we plot the ratio of the Compton-$y$ parameter
between the runs with and without AGN feedback at $z=0$. The left-hand
panel illustrates the case where bubbles are purely thermal, while the
central ($\alpha=2.4$) and right-hand panels ($\alpha=2.1$) refer to the
runs with CR bubbles. The Compton-$y$ parameter is essentially
determined by the line-of-sight integral of the thermal pressure. Since
AGN-driven bubbles modify the thermal pressure distribution within the
cluster, we observe variations of the estimated Compton-$y$
parameter. More specifically, the Compton-$y$ parameter is significantly
reduced in the central cluster regions for all three runs with AGN
feedback. As a side effect of solving the overcooling problem, the
self-regulated AGN feedback results in a lower central density that is
in better agreement with the density profiles derived from X-ray
measurements. In the cluster outskirts, however, the thermal pressure
is somewhat increased in the runs with AGN feedback. This implies that
the gas pressure distribution is somewhat ``puffed up'' in the cluster
outskirts, which compensates for central gas depletion. Given that the
thermal energy of the cluster is roughly constant and independent of the
presence or absence of AGN heating, the area integrated Compton-$y$
parameter turns out to be quite robust and is not changed much. It is
primarily a measure of the cluster's gravitational potential.

Still, this balance is not perfect as can be deduced by calculating the
total Compton-$y$ parameter $Y$ within $R_{\rm vir}$. In the case of
thermal bubbles, we find that the integrated Compton-$y$ parameter is
increased by $\sim 10\%$, while in the case of CR bubbles with spectral
index of $2.4$ this number is somewhat smaller, being of order
$7\%$. Interestingly, in the case of CR bubbles with a flatter spectral
slope the total Compton-$y$ parameter is instead lowered by $\sim 7\%$,
as a result of a substantial reduction of the thermal pressure in the
innermost regions. These results indicate that the integrated
Comptonization is sensitive at a level of $\sim 10\%$ to the nature of
AGN-driven bubbles and to the adopted feedback efficiency. This suggests
that the presence of AGN feedback contributes to the scatter in the
scaling relation between the cluster mass and $Y$, and the magnitude of
this effect depends on the detailed physics of the AGN feedback. Further
studies are needed to characterize the functional form of the scatter
and to quantify a possible mass dependence of this effect. It is clear
that the intended use of $Y$-measurements for high precision cosmology
with galaxy clusters will require an accurate understanding of the
detailed physical properties of AGN-inflated bubbles.

\section{Discussion}  \label{Discussion} 

In this study we have investigated a model for self-regulated BH
feedback with non-thermal relativistic particles in AGN-inflated
bubbles. To describe the cosmic ray particles, we have adopted the
formalism developed by \citet{Ensslin2007} and \citet{Jubelgas2007} for
the treatment of CR protons, and combined it with prescriptions for BH
growth and feedback, outlined in \citet{Springel2005b} and
\citet{Sijacki2007}. This allowed us to study the influence of CR
bubbles on their host clusters in self-consistent cosmological
simulations of galaxy cluster formation, and in particular, to compare
with the case where the bubbles are filled with purely thermal gas. Our
methods also allowed us to gain some insight on how strongly the bubble
morphologies and their propagation through the cluster atmosphere are
affected by a cluster's dynamical state and the gaseous bulk motions in
the ICM.

We have found that simulations with CR bubbles much more faithfully
resemble observational findings compared with the results for thermal
bubbles. This is mostly driven by the softer equation of state of gas
that is supported by non-thermal pressure, and by the longer dissipation
timescale of the CR component relative to thermal cooling. As a result,
CR bubbles can rise to the cluster outskirts and even leak into the
surrounding intergalactic medium. Still, their pressure support becomes
gradually less important as they move away from cluster central
regions. We have also found that the bubble morphologies in forming
clusters are rather complex, especially during major merger events,
where bubbles can become significantly perturbed from their initially
spherical shape and are displaced from their initial injection
axis. This implies that based on the spatial position of bubbles alone
it is not readily possible to infer whether the AGN jet has changed its
orientation between two successive outbursts or whether bulk motions in
the ICM have displaced buoyantly rising bubbles.

Interestingly, we have found in our cosmological simulations that
neither the BHARs nor the SFRs are very sensitive to the nature of the
bubble feedback. However, the temperature distribution in the ICM is
affected noticeably, given that CR bubbles provide non-thermal pressure
support in central cluster regions. Above all, this changes the galaxy
cluster temperature profiles which decline towards the innermost
regions, being in much better agreement with the observed temperature
profiles of cool core clusters. Correspondingly, the thermal pressure
distribution is altered as well by CR bubbles, with a reduction of the
thermal pressure in the centre and an increase in the outskirts. This
redistribution of the thermal pressure can modify the integrated
Compton-$y$ parameter at a level of $\sim 10\%$, an effect that we
expect to contribute to the scatter in the relationship between cluster
mass and $Y$.

We have evaluated how many significant bubble outburst the central BH in
our cluster is undergoing during the time interval from $z=2$ to $z=0$,
when the BH is in the low accretion, `radio-mode' regime. Based on this
we obtain an average duty cycle of $\sim 10^8\,{\rm yrs}$. We note
however that sudden inflows of gas towards the central BH triggered by
merging activity -- especially at $z=0.6$, which corresponds to the last
major merger of the host cluster -- can cause much more frequent bubble
events than the average value during a limited period. There is hence
considerable variability in the duty cycle related to the merger history
of the host halo, an issue we plan to investigate in more detail in a
forthcoming study.

In our simulations, the sizes of the bubbles are not determined by
calculating the jet physics from first principles, given that we cannot
resolve the initial stage of jet formation and bubble inflation by a jet
even in the most advanced state-of-art cosmological
simulations. Instead, we rely on simple physical parameterizations and
observational guidance to inject the bubbles with plausible sizes.
Certainly this is an important limitation of our work and may in part be
responsible for the rather large bubble sizes of up to a few hundred kpc
that we find in some cases. On the other hand, there are two additional
factors that may favor rather large bubble dimensions in our current
modelling. One is that we have kept the injection axis constant with
time. In some cases, the bubble injection time matches the buoyant rise
time in such a way that fresh CR protons are injected into an already
existing bubbles, making them reach a larger size. If this is also
realized in nature, it is potentially ambiguous whether large bubbles
are created by particularly powerful outbursts or result from several
smaller outbursts which combine together. 

 However, we note that if CR
  diffusion is efficient (which we have not considered in this study)
  it would be less probable that several bubble outbursts mix together into a
  large X--ray cavity. In this case an old bubble would already disperse
  before a new one could have time to reach it and combine with it. CR
  diffusion would also reduce the probability of finding a bubble at large
  enough distance from the cluster centre (and would also possibly modify the
  heating rates), thus more sophisticated studies and more stringent
  observational constrains are needed to pin-down the relevance of CR
  diffusion out of the bubbles.

The second factor which may bias the sizes of our simulated CR bubbles
high is the fact that observationally only relativistic electrons have
been detected so far in radio lobes, while we have conjectured in our
simulations that there should be a spatially extended distribution of CR
protons as well. Many of our CR bubbles, especially the ones that are
further away from the central AGN, are expected to have a rather aged
population of CR electrons, which would not have detectable radio
emission. In our picture these bubbles correspond to the so-called ghost
cavities which are identified as depressions in X--ray emission without
any notable radio emission, or only radio emission at lower frequencies.
Because of the very low contrast in projected surface brightness maps,
many of these ghost cavities are missed both in observations and
simulations. It will be an important task for future work to better
quantify this bias, and to find powerful observational constraints for
the relative content of relativistic electrons and protons in
AGN-inflated bubbles. Our simulations certainly suggest that CRs may
play a crucial role in shaping the central ICM properties and in
regulating AGN activity in clusters of galaxies.

\section*{Acknowledgements} We are grateful to Simon White and Lars Hernquist
for very helpful discussions and comments on this work. DS acknowledges the PhD fellowship of the
International Max Planck Research School in Astrophysics, and a
Postdoctoral Fellowship from the UK Science and Technology Funding
Council (STFC).

\bibliographystyle{mnras}

\bibliography{paper}

\begin{thebibliography}{60}
\expandafter\ifx\csname natexlab\endcsname\relax\def\natexlab#1{#1}\fi

\bibitem[{Allen} et~al.(2006){Allen}, {Dunn}, {Fabian}, {Taylor} \&
  {Reynolds}]{Allen2006}
{Allen} S.~W., {Dunn} R.~J.~H., {Fabian} A.~C., {Taylor} G.~B., {Reynolds}
  C.~S., 2006, \mnras, 372, 21

\bibitem[{B{\^i}rzan} et~al.(2004){B{\^i}rzan}, {Rafferty}, {McNamara}, {Wise}
  \& {Nulsen}]{Birzan2004}
{B{\^i}rzan} L., {Rafferty} D.~A., {McNamara} B.~R., {Wise} M.~W., {Nulsen}
  P.~E.~J., 2004, \apj, 607, 800

\bibitem[{Blanton} et~al.(2001){Blanton}, {Sarazin}, {McNamara} \&
  {Wise}]{Blanton2001}
{Blanton} E.~L., {Sarazin} C.~L., {McNamara} B.~R., {Wise} M.~W., 2001, \apjl,
  558, L15

\bibitem[{Boehringer} \& {Morfill}(1988)]{Boehringer1988}
{Boehringer} H., {Morfill} G.~E., 1988, \apj, 330, 609

\bibitem[{Bondi}(1952)]{Bondi1952}
{Bondi} H., 1952, \mnras, 112, 195

\bibitem[{Bondi} \& {Hoyle}(1944)]{Bondi1944}
{Bondi} H., {Hoyle} F., 1944, \mnras, 104, 273

\bibitem[{Br{\"u}ggen} \& {Kaiser}(2002)]{Brueggen2002}
{Br{\"u}ggen} M., {Kaiser} C.~R., 2002, \nat, 418, 301

\bibitem[{Chandran} \& {Dennis}(2006)]{Chandran2006}
{Chandran} B.~D., {Dennis} T.~J., 2006, \apj, 642, 140

\bibitem[{Churazov} et~al.(2001){Churazov}, {Br{\"u}ggen}, {Kaiser},
  {B{\"o}hringer} \& {Forman}]{Churazov2001}
{Churazov} E., {Br{\"u}ggen} M., {Kaiser} C.~R., {B{\"o}hringer} H., {Forman}
  W., 2001, \apj, 554, 261

\bibitem[{Churazov} et~al.(2007){Churazov}, {Forman}, {Vikhlinin}, {Tremaine},
  {Gerhard} \& {Jones}]{Churazov2007}
{Churazov} E., {Forman} W., {Vikhlinin} A., {Tremaine} S., {Gerhard} O.,
  {Jones} C., 2007, MNRAS submitted, astro-ph/0711.4686

\bibitem[{Clarke} et~al.(2005){Clarke}, {Sarazin}, {Blanton}, {Neumann} \&
  {Kassim}]{Clarke2005}
{Clarke} T.~E., {Sarazin} C.~L., {Blanton} E.~L., {Neumann} D.~M., {Kassim}
  N.~E., 2005, \apj, 625, 748

\bibitem[{Dalla Vecchia} et~al.(2004){Dalla Vecchia}, {Bower}, {Theuns},
  {Balogh}, {Mazzotta} \& {Frenk}]{DallaVecchia2004}
{Dalla Vecchia} C., {Bower} R.~G., {Theuns} T., {Balogh} M.~L., {Mazzotta} P.,
  {Frenk} C.~S., 2004, \mnras, 355, 995

\bibitem[{Di Matteo} et~al.(2005){Di Matteo}, {Springel} \&
  {Hernquist}]{DiMatteo2005}
{Di Matteo} T., {Springel} V., {Hernquist} L., 2005, \nat, 433, 604

\bibitem[{Dolag}(2004)]{Dolag2004}
{Dolag} K., 2004, in { The Riddle of Cooling Flows in Galaxies and Clusters of
  galaxies\/}, edited by T.~H. {Reiprich}, J.~C. {Kempner}, N.~{Soker},
  27--30, to be published electronically at
  http://www.astro.virginia.edu/coolflow/

\bibitem[{Dunn} et~al.(2005){Dunn}, {Fabian} \& {Taylor}]{Dunn2005}
{Dunn} R.~J.~H., {Fabian} A.~C., {Taylor} G.~B., 2005, \mnras, 364, 1343

\bibitem[{En{\ss}lin} et~al.(2007){En{\ss}lin}, {Pfrommer}, {Springel} \&
  {Jubelgas}]{Ensslin2007}
{En{\ss}lin} T.~A., {Pfrommer} C., {Springel} V., {Jubelgas} M., 2007, \aap,
  473, 41

\bibitem[{Fabian} et~al.(2006){Fabian}, {Sanders}, {Taylor}, {Allen},
  {Crawford}, {Johnstone} \& {Iwasawa}]{Fabian2006}
{Fabian} A.~C., {Sanders} J.~S., {Taylor} G.~B., {Allen} S.~W., {Crawford}
  C.~S., {Johnstone} R.~M., {Iwasawa} K., 2006, \mnras, 366, 417

\bibitem[{Forman} et~al.(2007)]{Forman2006}
{Forman} W., et~al., 2007, \apj, 665, 1057

\bibitem[{Gould}(1972)]{Gould1972}
{Gould} R.~J., 1972, Physica, 58, 379

\bibitem[{Guo} \& {Oh}(2008)]{Guo2007}
{Guo} F., {Oh} S.~P., 2008, \mnras, 384, 251

\bibitem[{Hoyle} \& {Lyttleton}(1939)]{Hoyle1939}
{Hoyle} F., {Lyttleton} R.~A., 1939, in { Proceedings of the Cambridge
  Philisophical Society\/}, vol.~35 of { Proceedings of the Cambridge
  Philisophical Society\/},  405--+

\bibitem[{Jenkins} et~al.(2001){Jenkins}, {Frenk}, {White}, {Colberg}, {Cole},
  {Evrard}, {Couchman} \& {Yoshida}]{Jenkins2001}
{Jenkins} A., {Frenk} C.~S., {White} S.~D.~M., {Colberg} J.~M., {Cole} S.,
  {Evrard} A.~E., {Couchman} H.~M.~P., {Yoshida} N., 2001, \mnras, 321, 372

\bibitem[{Jubelgas} et~al.(2008){Jubelgas}, {Springel}, {En{\ss}lin} \&
  {Pfrommer}]{Jubelgas2007}
{Jubelgas} M., {Springel} V., {En{\ss}lin} T., {Pfrommer} C., 2008, \aap, 481,
  33

\bibitem[{Katz} et~al.(1996){Katz}, {Weinberg} \& {Hernquist}]{Katz1996}
{Katz} N., {Weinberg} D.~H., {Hernquist} L., 1996, \apjs, 105, 19

\bibitem[{Lin} et~al.(2003){Lin}, {Mohr} \& {Stanford}]{Lin2003}
{Lin} Y.-T., {Mohr} J.~J., {Stanford} S.~A., 2003, \apj, 591, 749

\bibitem[{Loewenstein} et~al.(1991){Loewenstein}, {Zweibel} \&
  {Begelman}]{Loewenstein1991}
{Loewenstein} M., {Zweibel} E.~G., {Begelman} M.~C., 1991, \apj, 377, 392

\bibitem[{Mathews} \& {Brighenti}(2007)]{Mathews2007}
{Mathews} W.~G., {Brighenti} F., 2007, \apj, 660, 1137

\bibitem[{Mazzotta} et~al.(2002){Mazzotta}, {Kaastra}, {Paerels}, {Ferrigno},
  {Colafrancesco}, {Mewe} \& {Forman}]{Mazzotta2002}
{Mazzotta} P., {Kaastra} J.~S., {Paerels} F.~B., {Ferrigno} C., {Colafrancesco}
  S., {Mewe} R., {Forman} W.~R., 2002, \apjl, 567, L37

\bibitem[{McNamara} et~al.(2005){McNamara}, {Nulsen}, {Wise}, {Rafferty},
  {Carilli}, {Sarazin} \& {Blanton}]{McNamara2005}
{McNamara} B.~R., {Nulsen} P.~E.~J., {Wise} M.~W., {Rafferty} D.~A., {Carilli}
  C., {Sarazin} C.~L., {Blanton} E.~L., 2005, \nat, 433, 45

\bibitem[{Miniati}(2002)]{Miniati2002}
{Miniati} F., 2002, \mnras, 337, 199

\bibitem[{Miniati} et~al.(2001){Miniati}, {Ryu}, {Kang} \&
  {Jones}]{Miniati2001}
{Miniati} F., {Ryu} D., {Kang} H., {Jones} T.~W., 2001, \apj, 559, 59

\bibitem[{Navarro} et~al.(1996){Navarro}, {Frenk} \& {White}]{Navarro1996}
{Navarro} J.~F., {Frenk} C.~S., {White} S.~D.~M., 1996, \apj, 462, 563

\bibitem[{Navarro} et~al.(1997){Navarro}, {Frenk} \& {White}]{Navarro1997}
{Navarro} J.~F., {Frenk} C.~S., {White} S.~D.~M., 1997, \apj, 490, 493

\bibitem[{Nulsen} et~al.(2005){Nulsen}, {McNamara}, {Wise} \&
  {David}]{Nulsen2005}
{Nulsen} P.~E.~J., {McNamara} B.~R., {Wise} M.~W., {David} L.~P., 2005, \apj,
  628, 629

\bibitem[{Omma} et~al.(2004){Omma}, {Binney}, {Bryan} \& {Slyz}]{Omma2004a}
{Omma} H., {Binney} J., {Bryan} G., {Slyz} A., 2004, \mnras, 348, 1105

\bibitem[{Owen} et~al.(2000){Owen}, {Eilek} \& {Kassim}]{Owen2000}
{Owen} F.~N., {Eilek} J.~A., {Kassim} N.~E., 2000, \apj, 543, 611

\bibitem[{Pfrommer}(2008)]{Pfrommer2007c}
{Pfrommer} C., 2008, \mnras, 385, 1242

\bibitem[{Pfrommer} et~al.(2005){Pfrommer}, {En{\ss}lin} \&
  {Sarazin}]{Pfrommer2005}
{Pfrommer} C., {En{\ss}lin} T.~A., {Sarazin} C.~L., 2005, \aap, 430, 799

\bibitem[{Pfrommer} et~al.(2008){Pfrommer}, {En{\ss}lin} \&
  {Springel}]{Pfrommer2007b}
{Pfrommer} C., {En{\ss}lin} T.~A., {Springel} V., 2008, \mnras, 385, 1211

\bibitem[{Pfrommer} et~al.(2007){Pfrommer}, {En{\ss}lin}, {Springel},
  {Jubelgas} \& {Dolag}]{Pfrommer2007a}
{Pfrommer} C., {En{\ss}lin} T.~A., {Springel} V., {Jubelgas} M., {Dolag} K.,
  2007, \mnras, 378, 385

\bibitem[{Pfrommer} et~al.(2006){Pfrommer}, {Springel}, {En{\ss}lin} \&
  {Jubelgas}]{Pfrommer2006}
{Pfrommer} C., {Springel} V., {En{\ss}lin} T.~A., {Jubelgas} M., 2006, \mnras,
  367, 113

\bibitem[{Quilis} et~al.(2001){Quilis}, {Bower} \& {Balogh}]{Quilis2001}
{Quilis} V., {Bower} R.~G., {Balogh} M.~L., 2001, \mnras, 328, 1091

\bibitem[{Ruszkowski} \& {Begelman}(2002)]{Ruszkowski2002}
{Ruszkowski} M., {Begelman} M.~C., 2002, \apj, 581, 223

\bibitem[{Ruszkowski} et~al.(2008){Ruszkowski}, {En{\ss}lin}, {Br{\"u}ggen},
  {Begelman} \& {Churazov}]{Ruszkowski2007}
{Ruszkowski} M., {En{\ss}lin} T.~A., {Br{\"u}ggen} M., {Begelman} M.~C.,
  {Churazov} E., 2008, \mnras, 383, 1359

\bibitem[{Ryu} \& {Kang}(2004)]{Ryu2004}
{Ryu} D., {Kang} H., 2004, Journal of Korean Astronomical Society, 37, 477

\bibitem[{Ryu} et~al.(2003){Ryu}, {Kang}, {Hallman} \& {Jones}]{Ryu2003}
{Ryu} D., {Kang} H., {Hallman} E., {Jones} T.~W., 2003, \apj, 593, 599

\bibitem[{Sanders} \& {Fabian}(2007)]{Sanders2007}
{Sanders} J.~S., {Fabian} A.~C., 2007, \mnras, 381, 1381

\bibitem[{Sanderson} et~al.(2006){Sanderson}, {Ponman} \&
  {O'Sullivan}]{Sanderson2006}
{Sanderson} A.~J.~R., {Ponman} T.~J., {O'Sullivan} E., 2006, \mnras, 372, 1496

\bibitem[{Schmidt} et~al.(2002){Schmidt}, {Fabian} \& {Sanders}]{Schmidt2002}
{Schmidt} R.~W., {Fabian} A.~C., {Sanders} J.~S., 2002, \mnras, 337, 71

\bibitem[{Sijacki} \& {Springel}(2006{\natexlab{a}})]{Sijacki2006a}
{Sijacki} D., {Springel} V., 2006{\natexlab{a}}, \mnras, 366, 397

\bibitem[{Sijacki} \& {Springel}(2006{\natexlab{b}})]{Sijacki2006b}
{Sijacki} D., {Springel} V., 2006{\natexlab{b}}, \mnras, 371, 1025

\bibitem[{Sijacki} et~al.(2007){Sijacki}, {Springel}, {di Matteo} \&
  {Hernquist}]{Sijacki2007}
{Sijacki} D., {Springel} V., {di Matteo} T., {Hernquist} L., 2007, \mnras, 380,
  877

\bibitem[{Simionescu} et~al.(2008){Simionescu}, {Werner}, {Finoguenov},
  {B{\"o}hringer} \& {Br{\"u}ggen}]{Simionescu2007}
{Simionescu} A., {Werner} N., {Finoguenov} A., {B{\"o}hringer} H.,
  {Br{\"u}ggen} M., 2008, \aap, 482, 97

\bibitem[{Springel}(2005)]{Gadget2}
{Springel} V., 2005, \mnras, 364, 1105

\bibitem[{Springel} et~al.(2005){Springel}, {Di Matteo} \&
  {Hernquist}]{Springel2005b}
{Springel} V., {Di Matteo} T., {Hernquist} L., 2005, \mnras, 361, 776

\bibitem[{Springel} \& {Hernquist}(2002)]{SH2002}
{Springel} V., {Hernquist} L., 2002, \mnras, 333, 649

\bibitem[{Springel} \& {Hernquist}(2003)]{S&H2003}
{Springel} V., {Hernquist} L., 2003, \mnras, 339, 289

\bibitem[{Springel} et~al.(2001){Springel}, {Yoshida} \& {White}]{Springel2001}
{Springel} V., {Yoshida} N., {White} S.~D.~M., 2001, New Astronomy, 6, 79

\bibitem[{Tormen} et~al.(1997){Tormen}, {Bouchet} \& {White}]{Tormen1997}
{Tormen} G., {Bouchet} F.~R., {White} S.~D.~M., 1997, \mnras, 286, 865

\bibitem[{Yoshida} et~al.(2001){Yoshida}, {Sheth} \& {Diaferio}]{Yoshida2001}
{Yoshida} N., {Sheth} R.~K., {Diaferio} A., 2001, \mnras, 328, 669

\end{thebibliography}

\end{document}